\documentclass[reprint,showpacs,showkeys,superscriptaddress,preprintnumbers,amsmath,amssymb,floatfix,aps,prl]{revtex4-1}
\usepackage[utf8]{inputenc}
\usepackage[T1]{fontenc}

\usepackage{graphicx}
\usepackage{color}
\usepackage{amsfonts}
\usepackage{amsmath}
\usepackage{graphicx}
\usepackage{bm}
\usepackage{amssymb}
\usepackage{times}
\usepackage{dcolumn}
\usepackage{cases}
\usepackage{booktabs}
\usepackage{setspace}
\usepackage{array}
\usepackage{blindtext}
\usepackage{placeins}
\usepackage{physics}
\usepackage{hhline}

\makeatletter

\makeatletter 
\renewcommand{\fnum@figure}{\textbf{Figure~\thefigure}}
\makeatother

\begin{document}
\title{Ultra-Stable Weyl Topology Driven by Magnetic Textures in the Shandite Compound Co$_3$Sn$_2$S$_{(2-x)}$Se$_x$}

\author{Dang Khoa Le}
\email[E-mail: ]{kdl222@usf.edu}
\affiliation{Department of Physics, University of South Florida, Tampa, Florida 33620, USA}
\author{Eklavya Thareja}
\email[E-mail: ]{ethareja@usf.edu}
\affiliation{Department of Physics, University of South Florida, Tampa, Florida 33620, USA}
\author{Bektur Konushbaev}
\affiliation{Department of Physics, University of South Florida, Tampa, Florida 33620, USA}
\author{Gina Pantano}
\affiliation{Department of Physics, University of South Florida, Tampa, Florida 33620, USA}
\author{Tom Saunderson}
\affiliation{Peter Gr\"{u}nberg Institut and Institute for Advanced Simulation, Forschungszentrum J\"{u}lich and JARA, 52425 J\"{u}lich, Germany}

\author{Manh-Huong Phan}
\affiliation{Department of Physics, University of South Florida, Tampa, Florida 33620, USA}
\affiliation{Center for Materials Innovation and Technology, VinUniversity, Hanoi 100000, Vietnam}
\author{Yuriy Mokrousov}
\affiliation{Peter Gr\"{u}nberg Institut and Institute for Advanced Simulation, Forschungszentrum J\"{u}lich and JARA, 52425 J\"{u}lich, Germany}
\affiliation{Institute of Physics, Johannes Gutenberg University Mainz, 55099 Mainz, Germany}

\author{Jacob Gayles}
\affiliation{Department of Physics, University of South Florida, Tampa, Florida 33620, USA}

\begin{abstract}
We employ state-of-the-art first-principles calculations to investigate the shandite compounds Co$_3$Sn$_2$S$_2$, Co$_3$Sn$_2$SeS, and Co$_3$Sn$_2$Se$_2$, which host Weyl fermions and complex magnetic textures. Their magnetic structures are governed primarily by exchange interactions and magnetocrystalline anisotropy, whereas the symmetry-allowed alternating-layer Dzyaloshinskii--Moriya interaction (DMI) is found to be negligible. We identify a previously unrecognized spin-chiral interaction (SCI) arising from the kagome lattice topology, which plays a decisive role in stabilizing the experimentally observed magnetic textures. The extracted magnetic parameters reproduce experimental trends, with the SCI emerging as a novel and dominant contribution. The calculated SCI strengths are 0.78 meV, 0.86 meV, and 0.87 meV for Co$_3$Sn$_2$S$_2$, Co$_3$Sn$_2$SeS, and Co$_3$Sn$_2$Se$_2$, respectively. Furthermore, we demonstrate that short-wavelength magnetic textures drive phase transitions of the Weyl nodes, resulting in band flattening and the opening of an emergent gap. This newly identified SCI, together with the associated electronically driven phase transitions, provides a promising route for manipulating transport properties in spintronic devices.

\end{abstract}

\maketitle


The cornerstone of modern electronics rests on the ability to switch conductivity states through external fields, such as n-p junctions and gate transistors, or by altering spin alignment, as seen in giant magnetoresistance devices. Modern devices demand a deeper understanding of how to realize high mobility and controllable coupling between electronic and magnetic structures. Topological materials, via spin--momentum--locked bands, require the lack of time reversal or inversion symmetry and pave a way for paradigm shifts in electronic and magnetic phenomena~\cite{BurkovRev2018,Bernevig2022}. Topological materials, including Dirac, Weyl (magnetic and nonmagnetic), and nodal-line semimetals, exhibit remarkable features resulting in large Berry curvature effects such as the Anomalous Hall effect (AHE), suppressed backscattering, and high carrier mobilities~\cite{BurkovRev2018,TopmatYan2017,Bernevig2022}. Despite these advances, systematic manipulation of Weyl and Dirac node crossings has so far been achieved primarily in Moir\'{e} bilayer graphene~\cite{Ohta2006,Cao2018,Andrei2020} and charge-density-wave systems~\cite{Gooth2019,Shi2021}. In magnetic systems, the interplay of Berry curvatures defined in distinct geometric spaces hinders progress, as it is not yet well understood~\cite{hanke2017mixed,Niu2019}. The antisymmetric Dzyaloshinskii--Moriya interaction (DMI) ~\cite{Dzyaloshinsky1958,Moriya1960}, which requires broken inversion symmetry and is closely tied to phase-space Berry curvature~\cite{freimuth2013phase}, appears to be incompatible with magnetic Weyl semimetals, which typically preserve inversion symmetry in the momentum space Berry curvature.

Nominally, the DMI is the crucial chiral ingredient required to stabilize real-space topological spin textures such as skyrmions~\cite{bogdanov1989,Pfleiderer2009}, which have attracted strong interest for potential applications in racetrack memory devices~\cite{Parkin2008,Tomasello2014}. The primary electrical signature of skyrmions and their size is the topological Hall effect~\cite{Bruno2004}; however, in multi-band metals, this contribution is typically minimal compared to the anomalous and ordinary Hall effects~\cite{Kimbell2022}. 
In the simplest picture, the size of magnetic textures is determined by competing energy scales, where domain wall widths are set by the ratio of exchange to the square root of magnetocrystalline anisotropic energy, while skyrmion sizes scale with the ratio of DMI to exchange~\cite{heide2008dzyaloshinskii,wang2018theory}. Stabilizing compact magnetic textures requires either a large DMI or strong anisotropy terms that scale quadratically and linearly with spin-orbit coupling, respectively~\cite{heide2008dzyaloshinskii,jiang2025higher}. The DMI is both large and anisotropic, which can dramatically alter the spin-momentum locking in topological materials and diminish their mobility. However, in the presence of frustrated magnetic lattices, especially kagome lattice symmetry, a novel spin-chiral interaction (SCI) emerges---which is intrinsically maximized by the local geometry~\cite{grytsiuk2020topological,singh2024higher,kolincio2023kagome,zhang2025valley}. The SCI can stabilize magnetic textures~\cite{grytsiuk2020topological} without sacrificing the Weyl topology and other novel transport phenomena\cite{yu2025quantum}.

\begin{figure*}[t]
    \includegraphics[width=\textwidth]{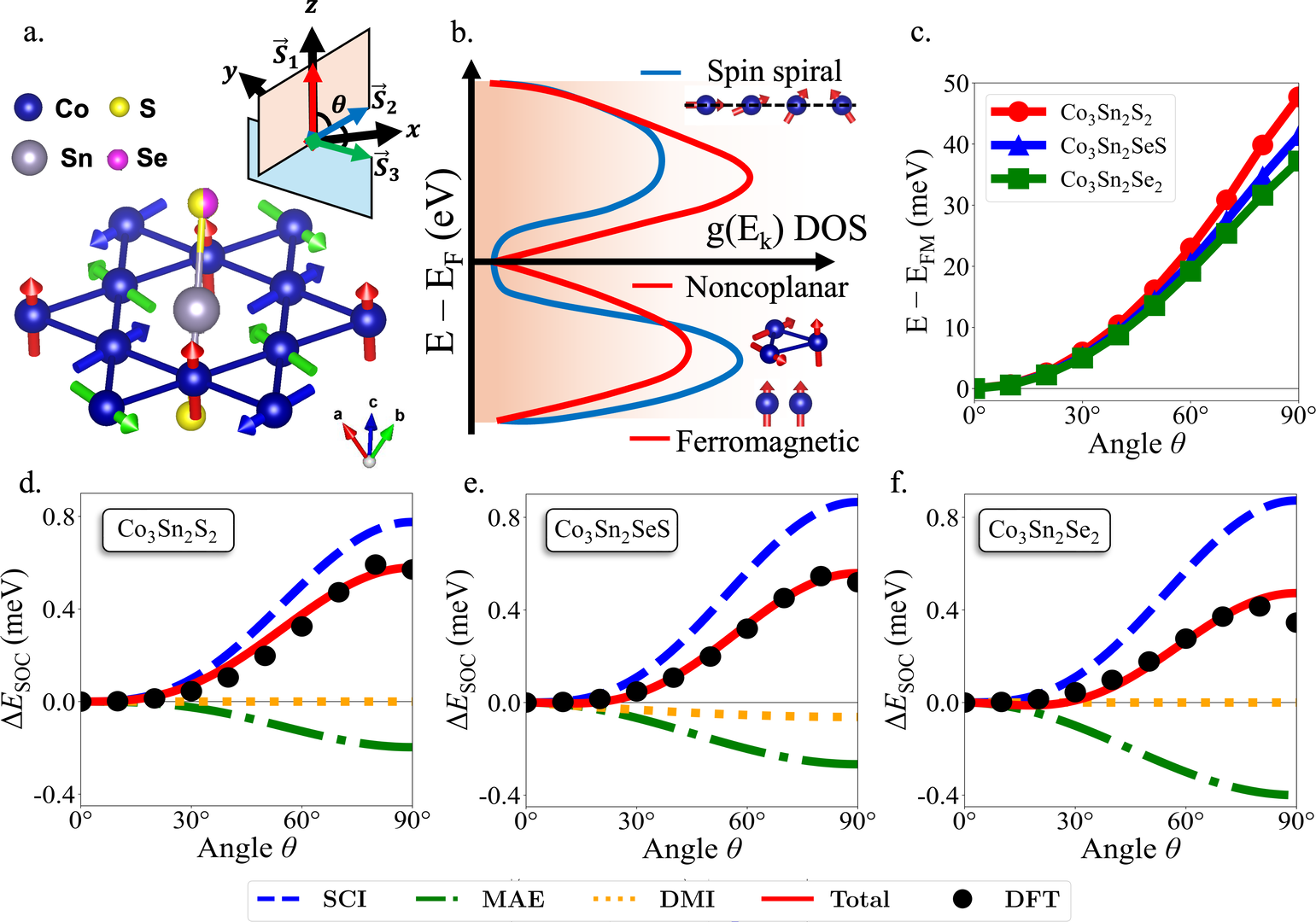}
\caption{(a) Side view of lattice structure of Co$_3$Sn$_2$S$_{(2-x)}$Se$_x$, where Co, Sn, S, and Se atoms are indicated in blue, gray, yellow, and green, respectively. (b) The kagome lattice formed by the Co atoms shows the noncoplanar magnetic state with a nonzero scalar chiral product. The inset displays the magnetic directions of the Co$_1$, Co$_2$, and Co$_3$ atoms in the Cartesian coordinate system, labeled with red, blue, and green arrows, respectively. (c) The energy dispersion relation of all shandite compounds with the noncoplanar spin orientations. (d--f) The comparison of spin-orbit-induced energy in (d) Co$_3$Sn$_2$S$_2$, (e) Co$_3$Sn$_2$SeS, and (f) Co$_3$Sn$_2$Se$_2$ between the DFT calculations (circles) and the spin-orbit contribution terms (lines), where the antisymmetric exchange (DMI) is shown as a green dash--dot line, while the spin-chiral interaction (SCI) is shown as a blue dashed line.}
    \label{FIG-1}
\end{figure*}
Such a case is realized in the layered shandite compound Co$_3$Sn$_2$S$_2$ and its stoichiometric variants Co$_3$Sn$_2$S$_{(2-x)}$Se$_x$, which crystallize in the rhombohedral space group $R\overline{3}m$ (No.,166). The magnetic Co atoms form a kagome network interlaced with Sn and S/Se atoms, with inversion symmetry preserved for $x=0,2$ but broken for $x=1$ (see Fig. 1 a). The symmetry allows for anti-aligned DMI~\cite{van2020layer} vectors on parallel kagome planes of Co atoms that globally sum to zero for $x=0,2$ and should nominally be finite for $x=1$, which lacks inversion symmetry. Co$_3$Sn$_2$S$_2$ was among the first predicted and experimentally confirmed magnetic Weyl semimetal, exhibiting an anomalous Hall conductivity ($\sigma_H^A = 1130$ $\Omega^{-1}$ $cm^{-1}$) close to the maximum value for a two-band system~\cite{liu2018giant,mitscherling2020longitudinal}. Together with their Se-substituted counterparts, these compounds display Curie temperatures ranging from 150~K to 180~K ~\cite{sakai2013magnetic,shin2021Tc} that decrease with increasing Se concentration. However, the precise magnetic ground state of Co$_3$Sn$_2$S$_{(2-x)}$Se$_x$ remains inconclusive.  Furthermore, the microscopic origins of the in-plane magnetic stabilization and anisotropy remain controversial~\cite{barua2025competing,guguchia2020tunable}. Reports include magnetic susceptibility measurements indicating an A-phase pocket reminiscent of skyrmion-hosting systems~\cite{wu2020observation,kassem2017low}, and large conductivity changes in the presence of domain wall networks~\cite{fujiwara2024giant,sugawara2019magnetic,pate2025anomalies}, quasi-flatbands associated with the magnon mode population~\cite{yin2019negative,nag2022correlation, xu2020electronic}. In this work, we resolve the outstanding questions related to the competing exchange terms and demonstrate how spin-chiral interaction (SCI) and different magnetic interactions influence the exotic, topology-driven bands that emerge in this shandite family.

We use first-principles calculations to investigate the magnetic structure and its influence on the topological electronic structure in the magnetic Weyl semimetals Co$_3$Sn$_2$S$_{(2-x)}$Se$_x$ for $x=0,1,2$.  We evaluate the SCI and DMI fields for all systems and observe that even in the inversion-asymmetric case ($x = 1$), the SCI is significantly larger than the DMI. In all cases of the SCI calculations with noncoplanar spin configuration, the Weyl topology is unperturbed. For high symmetry propagating spin-spirals, the nodes transition from Type-I Weyl to Type-II Weyl, and finally to a gapped state, as the wavelength is reduced to short periods. The Weyl nodes in the inversion-asymmetric compound Co$_3$Sn$_2$SSe ($x=1$) show the strongest stability against magnetic textures. We further calculate the exchange stiffness using homogeneous spin spirals, i.e., frozen magnons, along the major crystallographic axes, obtaining values that qualitatively agree with and follow the trends of previous experimental reports of the Curie temperature~\cite{sakai2013magnetic,shin2021Tc}. Our calculations of the magnetocrystalline anisotropy (MCA) reveal a systematic increase with enhanced spin-orbit interaction from $x = 0$ to $x = 2$.

Fig.\ref{FIG-1}. a illustrates noncoplanar spin moment arrangements that create a maximized scalar chirality product $\chi_{ijk}=\vec{S}_i \cdot (\vec{S}_j \times \vec{S}_k)$, which leads to nonzero SCI. Fig.\ref{FIG-1}. b shows the evolution of the projected density of states (DOS) from a conventional ferromagnetic Weyl and noncoplanar Weyl to a gapped noncollinear spin-spiral state. The angle $\theta$ (inset of Fig.\ref{FIG-1}. a)  describes the change of spin-configuration of each magnetic Co atom in noncoplanar configurations. Considering the dependence of energy dispersion on the angle $\theta$ without spin-orbit coupling (SOC), we calculated the exchange energy of Co$_3$Sn$_2$S$_{(2-x)}$Se$_x$. Co$_3$Sn$_2$S$_2$ exhibits the largest exchange energy, while Co$_3$Sn$_2$Se$_2$ shows the weakest. By including the SOC effect, the intra-cell energy dispersion of our shandite family depends on three magnetic interactions, the DMI, magnetocrystalline anisotropy (MAE), and the spin-chiral interaction (SCI). Using Eq. \ref{eq4} based on the spin configuration in Eq.\ref{eq2}, we obtained the energy functions as $E_{DMI} \sim \sin(\theta)$ (orange dashed lines), $E_{MAE} \sim \sin^2(\theta)$ (green dash-dot lines) and $E_{SCI} \sim \sin^3(\theta)$ (blue broken lines). We evaluated the contribution of these interactions based on first, second, and third order correction terms to the exchange energy. The DMI in the out-of-plane direction is nearly zero in comparison to the SCI, with the largest energy in the inversion lacking Co$_3$Sn$_2$SSe. The strength of SCI is $0.78$ meV, $0.86$ meV, and $0.87$ meV for $x=0,1$ and $2$, respectively, and in correspondence with the increase in SOC strength.

\begin{figure*}[t]
    \centering
    \includegraphics[width=\textwidth]{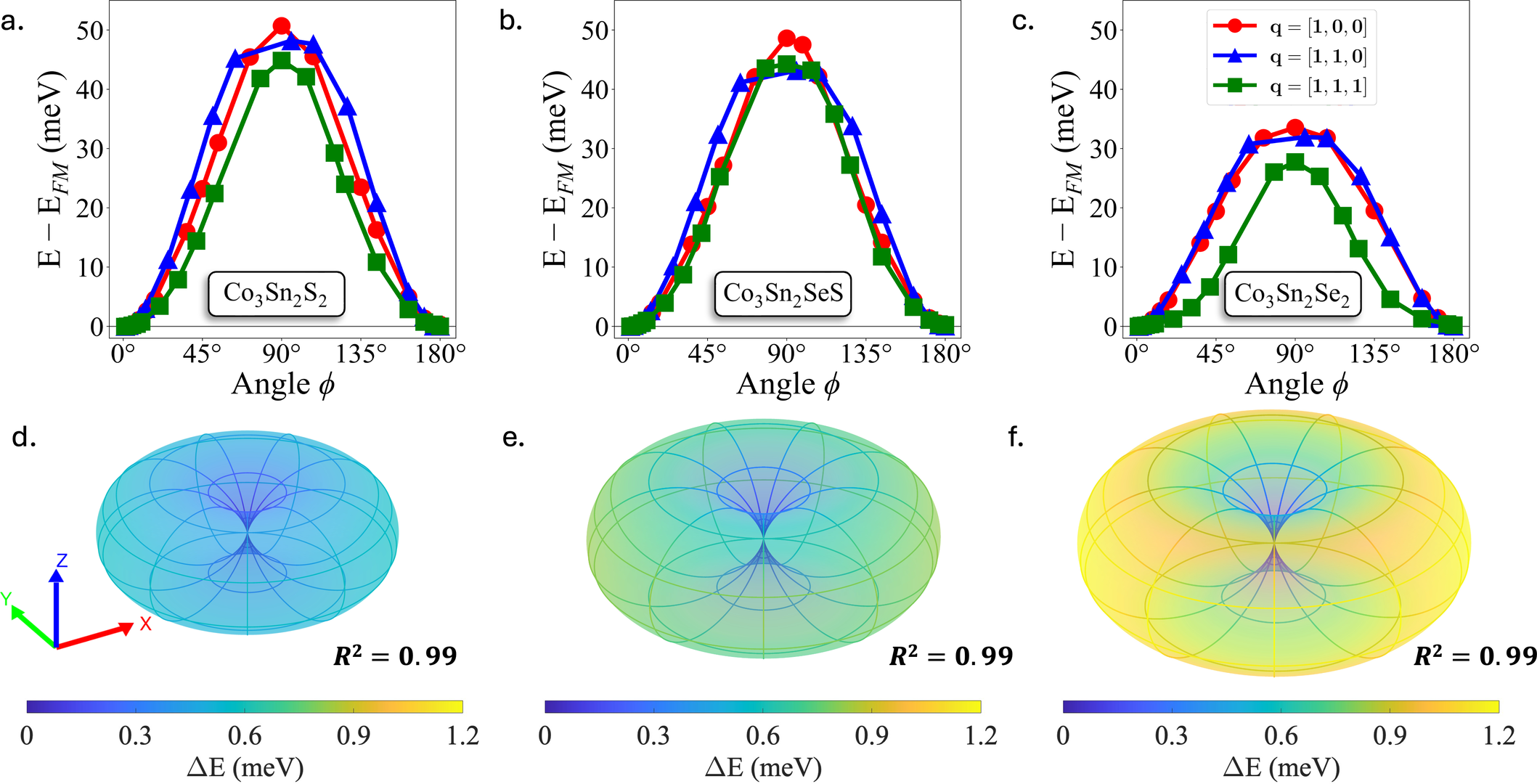}
    \caption{The dispersion relation in flat spin spirals and magnetocrystalline anisotropy surface of all shandite compounds. (a-c) Energy dispersion of spin spirals along [1,0,0] [1,1,0] and [1,1,1] directions (without SOC consideration), for (a) Co$_3$Sn$_2$S$_2$, (b) Co$_3$Sn$_2$SeS, and (c) Co$_3$Sn$_2$Se$_2$. (d-f) Three-dimensional magnetocrystalline anisotropy surfaces reflect the magnetocrystalline anisotropy energy of (d) Co$_3$Sn$_2$S$_2$, (e) Co$_3$Sn$_2$SeS, and (f) Co$_3$Sn$_2$Se$_2$.}
    \label{FIG-2}
\end{figure*}

We calculated the spin exchange stiffness ($A$) and magnetocrystalline anisotropy ($K_i$) of the shandite compounds. The exchange energy was calculated for two spin configurations, the homogeneous spin spiral and noncoplanar magnetic texture. The energies for our spin-spiral calculations were computed without SOC as flat spin spirals with a cone angle $\theta_i=\pi/2$. Fig. \ref{FIG-2}(a-c) shows the energy dispersions for the spin spirals along the $q=[1,0,0]$ (red-circles), $q=[1,1,0]$ (blue-triangles) and the $q=[1,1,1]$ (green-squares) directions in the three Co$_3$Sn$_2$S$_{(2-x)}$Se$_x$ compounds. $A$ is fitted to the energy dispersion of the spin spiral by the simplified version of  $E(q)=A|\mathbf{q}|^2$~\cite{heide2009describing}. Table \ref{Exchange_MAE} shows the calculated spin stiffness exchange for the noncoplanar 2Q and spin spiral magnetic textures. The largest $A$ for all compounds is along the $[1,1,0]$ direction, where  $A$  decreases from $450.88$\,meV for Co$_3$Sn$_2$S$_2$ ($x=0$) to $332.25$\,meV for Co$_3$Sn$_2$SeS ($x=1$) and $264.07$\,meV for Co$_3$Sn$_2$Se$_2$ ($x=2$). The Curie temperature was calculated from the spin exchange stiffness from $A=C^{2/3}kT_c$, where C is determined by intrinsic magnetic variables, including critical exponent $\beta$ and magnetization at absolute temperature $0K$~\cite{kuz2020exchange}. The Curie temperature of Co$_3$Sn$_2$S$_2$ is 173K, in good agreement with the experimental value of 169K. When the Se/S ratio increases, the Curie temperature decreases due to a decrease in spin-exchange stiffness and in line with experimental trend ~\cite{sakai2013magnetic,shin2021Tc}.

The stabilization of the magnetic structure is also determined by the magnetic anisotropy, which indicates the preferred alignment of the magnetic sites. The magnetic anisotropic energy of Co$_3$Sn$_2$S$_{(2-x)}$Se$_x$ is calculated in collinear magnetic textures with the presence of SOC. We plot the magnetic anisotropy energy surface of the Co$_3$Sn$_2$S$_2$, Co$_3$Sn$_2$SeS, and Co$_3$Sn$_2$Se$_2$ in Fig.\ref{FIG-2}(d-f). The MCA constants, $K_i$,  are fitted to the energy surface by the symmetry of the crystal surface~\cite{doring1957richtungsabhangigkeit}. Fig.\ref{FIG-2}(d-f) illustrates a perpendicular magnetic anisotropy for all compounds, where the MCA energies along the z-direction are the lowest when the magnetic spin prefers to point out of the Kagome lattice. These results are consistent with those observed experimentally ~\cite{shen2019anisotropies}. Among the three compounds, Co$_3$Sn$_2$Se$_2$ exhibits the highest MCA energy (1.2 meV), likely due to the large spin-orbit coupling of the Se atom compared to that of the S counterpart. This result is also consistent with the differences in the energy landscapes of the three systems shown in Fig.\ref{FIG-2}(d--f). Using the obtained spin exchange stiffness and anisotropy energy, the domain wall width can be estimated from the relation $\delta_{DW} \simeq \pi\sqrt{A_{ex}/K}$. The calculated domain wall width of the Co$_3$Sn$_2$S$_2$ shandite compound is $\delta_{DW}=3.47$ nm, whereas experimental measurements report $\delta_{DW} \sim 2$ nm~\cite{lee2022observation,shiogai2022electrical}. This discrepancy likely reflects the influence of complex magnetic interactions, including spin-chiral interactions. The presence of such an interaction in shandite systems can further reduce the domain wall width, promoting the stabilization of noncollinear magnetic textures.

\begin{table}[h]
\centering
\begin{tabular}{lccc}
\hhline{====} 
\textbf{Exchange stiffness} & \textbf{Co$_3$Sn$_2$S$_2$} & \textbf{Co$_3$Sn$_2$SeS} & \textbf{Co$_3$Sn$_2$Se$_2$} \\
\hline
\textbf{$A_{[1,0,0]}$} $(meV)$  & 339.08 & 299.00 & 270.92 \\
\textbf{$A_{[1,1,0]}$} $(meV)$  & 450.88 & 332.25 & 264.07 \\
\textbf{$A_{[1,1,1]}$} $(meV)$  & 284.50 & 277.24 & 151.20 \\
\textbf{$A_{SC}$} \; \; $(meV)$  & 210.53 & 186.74 & 174.08 \\
\hhline{====} 
\textbf{Anisotropic energy} & \textbf{Co$_3$Sn$_2$S$_2$} & \textbf{Co$_3$Sn$_2$SeS} & \textbf{Co$_3$Sn$_2$Se$_2$} \\
\hline
\textbf{$K_0$} $(meV)$  &  0.59  &  0.82  &  1.19  \\
\textbf{$K_1$} $(meV)$  & -0.60  & -0.83  & -1.20  \\
\textbf{$K_2$} $(meV)$  &  0.003 &  0.002 &  0.001 \\
\textbf{$K_3$} $(meV)$  &  0.001 & -0.001 & -0.023 \\
\hhline{====} 
\textbf{Spin-chiral} & \textbf{Co$_3$Sn$_2$S$_2$} & \textbf{Co$_3$Sn$_2$SeS} & \textbf{Co$_3$Sn$_2$Se$_2$} \\
\hline
\textbf{$SCI$} $(meV)$ & $0.78$ & $0.86$ & $0.87$  \\
\hhline{====} 
\textbf{DMI} & \textbf{Co$_3$Sn$_2$S$_2$} & \textbf{Co$_3$Sn$_2$SeS} & \textbf{Co$_3$Sn$_2$Se$_2$} \\
\hline
\textbf{$D$} $(meV)$ & $0.000$ & $-0.063$ & $0.000$  \\
\hhline{====} 
\end{tabular}
\caption{Computed exchange and magnetic anisotropic interactions for the propagation directions of the q vector in Co$_3$Sn$_2$S$_{(2-x)}$Se$_x$.}
\label{Exchange_MAE}
\end{table}

\begin{figure}[t]
    \centering
    \includegraphics[width=0.95\columnwidth]{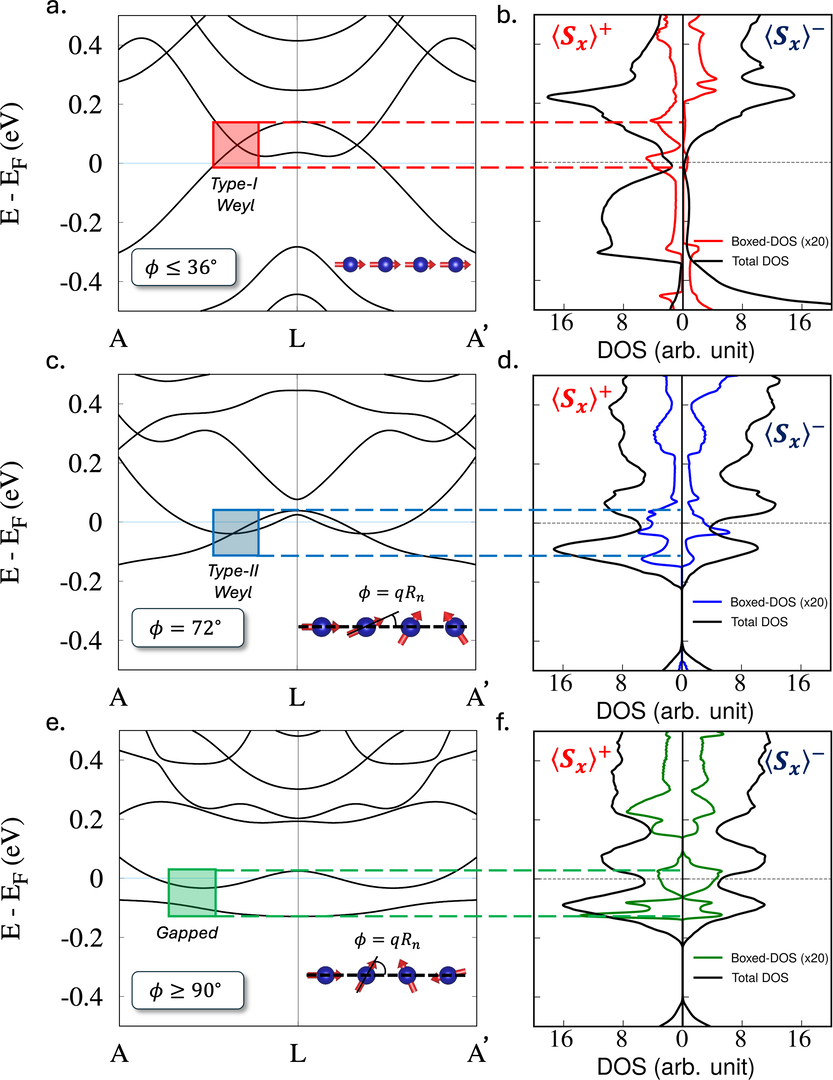}
    \caption{ Electronic band structures and density of states (DOS) of Co$_3$Sn$_2$S$_2$ with different homogeneous spin spiral angles $\phi$. Band structure and DOS are illustrated for $\phi=0 ^{\circ}$ in (a-b), $\phi=72^{\circ}$ in (c-d), and $\phi=108^{\circ}$ in (e-f), respectively. The spin projection in DOS is along the x-direction.}
    \label{FIG-3}
\end{figure}

Magnetic textures, such as domain walls, significantly modify the Weyl electronic band structure in Co$_3$Sn$_2$S$_{(2-x)}$Se$_x$ compounds. In Fig. \ref{FIG-3}(a,c,e), we show how varying the length of homogeneous flat spin spirals ($\theta_i = \pi/2$) alters the band structure near Weyl crossings in Co$_3$Sn$_2$S$_2$. In the ferromagnetic state, Co$_3$Sn$_2$S$_{2-x}$Se$_x$ hosts magnetic Weyl cones near the high-symmetry L point (where $L=[1/2;0;0]$). It is worth mentioning that these topological features do not appear along the high-symmetry k path. Therefore, we select a specific k-path that intersects the Weyl nodes. The positions in k-space of the Weyl nodes in the ferromagnetic ground state are $W_{S_2}=[0.313; -0.056; -0.056]$, $W_{SeS}=[0.284; -0.085; -0.085]$, and $W_{Se_2}=[0.276;-0.102;-0.102]$. Here, the coordinates of the above k-points are in the fraction coordinates of the first Brillouin zone. Applying rotational symmetry $C_{3z}$ yields similar Weyl nodes as reported in~\cite{xu2018topological}. In Fig. \ref{FIG-3}(a), Type-I Weyl crossings are indicated in the red box for small spin spiral angles ($\phi \leq 45$) in Co$_3$Sn$_2$S$_2$. These features originate from the absence of time-reversal symmetry in Co$_3$Sn$_2$S$_2$ and Co$_3$Sn$_2$Se$_2$. With the variation of $\phi$ values, i.e. $\phi=72^\circ$, the Weyl nodes of these shandite compounds undergo a transition from Type-I Weyl to Type-II Weyl, highlighted in the blue box in Fig.\ref{FIG-3}(c). When $\phi$ surpasses $90^\circ$, these Weyl nodes become gapped as seen in the green box of Fig.\ref{FIG-3}(e). The behavior of Weyl crossings in the band structures is presented in Fig.S1. However, due to inversion symmetry breaking arising from the substitution of S with Se atoms, we observe the magnetic Weyl nodes persist in Co$_3$Sn$_2$SeS even with $\phi \geq 90^\circ$. In particular, the electronic bands exhibit double crossings along the chosen $LA$ and $LA'$ k-path. Further investigations are necessary to understand the nature of these emergence crossings in Co$_3$Sn$_2$SeS compounds. Shandite systems undergo a sequence of topological phase transitions first, from Type-I to Type-II Weyl semimetals (characterized as a second-order transition), and finally to a gapped state (characterized as a first-order transition). Under the noncollinear spin orientation, the existence of Weyl crossings in the band structures originates from the vertical mirror symmetry $\sigma_v$, while rotational symmetry $C_{3z}$ preserves magnetic Weyl points with opposite Chern numbers $+1$ and $-1$ in the presence of SOC~\cite{xu2018topological}. 

The topological phase transitions are also reflected in the density of states (DOS) shown in Fig.\ref{FIG-3} (b,d,f). We show the total DOS of Co$_3$Sn$_2$S$_2$ and the DOS in the regions where the Weyl points are located. A box of k-mesh is constructed around the k-path $AA'$ to demonstrate the variation of the DOS, as shown by the corresponding colors as in band structures in Fig.~\ref{FIG-3}(b,d,f). We also projected DOS along the x-direction, where the left panel DOS corresponds to a positive projection, and the right panel DOS corresponds to a negative projection. As shown in Fig.\ref{FIG-3}.(b) for $\phi\leq 36^\circ$, the majority spin dominates the DOS near the Fermi energy, which is the positive region for spin-projected along the $x$ direction $\langle S_x \rangle$. However, when $\phi$ increases, there is a redistribution from positive $\langle S_x \rangle$ to negative $\langle S_x \rangle$. It is apparent in Fig.\ref{FIG-3}(d) there is an appearance of the DOS peaks near Fermi energy in the negative regions. The linear behavior appeared in the electronic states from the boxed DOS curve when the Weyl crossings gapped $(\phi \geq 90^\circ)$. Fig.S2 in Supplementary Information (SI) shows the band corresponding to the above DOS and reflects the gapped signature of the Weyl crossings. However, the behavior of DOS along the noted box that contains Weyl nodes does not affect the overall DOS of the systems. Since the Weyl cones only exist along a specific path $(AA')$ that passes through the high symmetry point L in the first Brillouin zone, not along the whole Brillouin zone. As the angle $\phi$ changes, the overall band structure of the shandite system alters drastically, while the magnetic Weyl crossings undergo a steady topological phase transition from type-I to type-II Weyl and eventually gapped as shown in Fig.\ref{FIG-3}. Similar behavior can be found in Co$_3$Sn$_2$Se$_2$. Nevertheless, the magnetic Weyl nodes in Co$_3$Sn$_2$SeS still persist even with $\phi \geq 90^\circ$ and form double crossings in the electronic band structure, leading to the nonzero DOS around the regions of the nodes.

\begin{figure}[t]
    \centering
    \includegraphics[width=0.95\columnwidth]{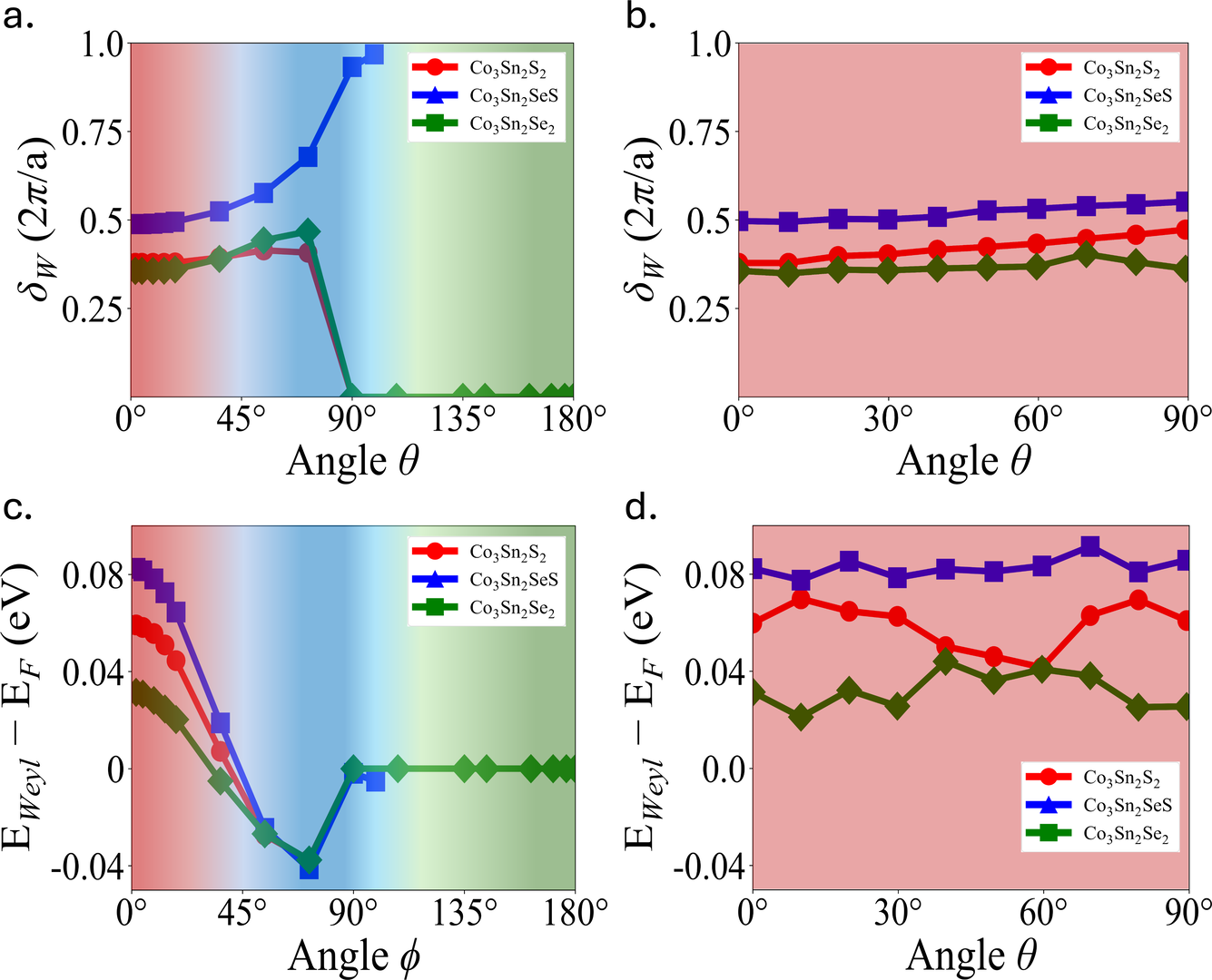}
    \caption{Topological phase transitions of Co$_3$Sn$_2$S$_{2-x}$Se$_x$ are revealed, where Co$_3$Sn$_2$S$_2$, Co$_3$Sn$_2$SeS, and Co$_3$Sn$_2$Se$_2$ are in red, blue and green lines, respectively. (a-b)  Distance between two Weyl crossings is plotted as a function of the spin spiral angle $\phi$ (along the direction q = [1, 0, 0]) and the noncoplanar angle $\theta$. (c-d) Position of the Weyl pairs according to the Fermi energy as a function of both the mentioned angles.}
    \label{FIG-4}
\end{figure}

To obtain a thorough understanding of the evolution of homogeneous spin spiral to Weyl cones along in both momentum space and its energy in the electronic band structures, Fig.\ref{FIG-4} displays the positions of these magnetic Weyl points in momentum space and their energy according to the Fermi energy. While Fig. \ref{FIG-4}(a) and (b) illustrate the separation between the two Weyl nodes in the shandite system, Fig. \ref{FIG-4}(c) and (d) show the positions of the Weyl crossings relative to the Fermi energy for all three compounds. We also indicate Type-I to Type-II and gapped regions in red-, blue-, and green-colored regions to demonstrate the topological phase transition mentioned in the previous section.
 
Under the noncollinear spin spiral, for Co$_3$Sn$_2$S$_2$ and Co$_3$Sn$_2$Se$_2$, the distance of Weyl-points increases slightly as systems transform from Type-I (red regions) to Type-II Weyls (blue regions), as shown in Fig.\ref{FIG-4}(a). During this process, we observed a downward shift of the Weyl crossing energies with respect to the Fermi energy, which is evident in Fig. \ref{FIG-4}(c). In addition, for Co$_3$Sn$_2$SeS, the Weyl-point distance increases, reaching nearly twice that of the ferromagnetic state. On the other hand, both Weyl cones of Co$_3$Sn$_2$S$_2$ and Co$_3$Sn$_2$Se$_2$ are gapped (indicated in the green region) as the spin spiral angle $\phi \geq 90^\circ$. When increasing $\phi$ even further, these Weyl crossings persist in a gapped state with $x=0,2$ in Co$_3$Sn$_2$S$_{(2-x)}$Se$_x$. However, for Co$_3$Sn$_2$SeS, the Weyl nodes persist as a function of the angle $\phi$. For the spin spiral direction $q=[1,1,1]$, we observe similar topological phase transitions, progressing from Type-I Weyl to Type-II Weyl, and ultimately to a gapped state. Nevertheless, the Weyl topological phase does not persist in the $q=[1,1,0]$ case, where the Weyl crossings become gapped with only a small change of spin spiral angle $\phi \geq 45^\circ$. These results are available in Fig. S3. For the noncoplanar case shown in Fig.\ref{FIG-4}(b,d), we found that minor variations in the separation of Weyl-crossings among the three systems, along with the shifts in their energy positions with varying angle $\theta$. This indicates that the Weyl topology remains Type-I Weyl semimetal of all three compounds during the noncoplanar spin configurations.

The SCI becomes the dominant interaction, likely to drive complex magnetic textures and enhance the stability of Weyl states in shandite systems. To evaluate the strength of SCI against spin-exchange stiffness, we take into account their ratio as follows SC/A. For our systems, that of Co$_3$Sn$_2$S$_2$, Co$_3$Sn$_2$SeS, and Co$_3$Sn$_2$Se$_2$ are S$_2 = 0.0031$, SeS$= 0.0037$, and Se$_2=0.0041$. When comparing the ratio of SC/A of our system to other calculated compounds, namely FeGe and MnGe~\cite{grytsiuk2020topological,gayles2015dzyaloshinskii}, the authors found that the SC/A ratios of FeGe and MnGe are FeGe$=0.0004$ and MnGe$=0.0113$, respectively, indicating that MnGe exhibits stronger SCI than FeGe.  In MnGe, the SCI has been suggested as the origin of novel and more complex spin textures~\cite{grytsiuk2020topological}, and a similar mechanism may also be at play in shandite compounds, which exhibit a significantly reduced exchange stiffness.

As we investigate the behavior Weyl nodes with the effect of SCI, Co$_3$Sn$_2$S$_{2-x}$Se$_x$ displays the Type-I Weyl topology. This highlights that the Weyl topology exhibits high stability under noncoplanar spin configurations. The tunability of Weyl nodes through control of the spin spiral angle directly influences the electronic and accompanying transport properties of the shandite materials. The Berry curvature is determined not only by the Weyl-node separation but also by the specific type of Weyl node, including type-I, tilted, and type-II, as reflected in the band structures. In all compounds considered, the momentum-space Weyl spin texture makes a negligible contribution to the DMI energy, indicating the absence of phase-space geometric terms. The variation of the spin spiral angle drives a phase transition from type-I Weyl to type-II Weyl, leading to enhanced Berry-curvature contributions arising from the Weyl-cone tilt. The emergence of the pronounced Berry curvature near the Fermi surface is expected to manifest experimentally as an increase in magnetoconductivity, arising from the broken time-reversal symmetry associated with the tilted Weyl spectrum. The presence of tilted and type-II Weyl nodes thus enhances conductivity, especially hall conductivity in the shandite systems~\cite{zyuzin2016intrinsic, burkov2011weyl, das2019linear}. When the Weyl crossings evolve from type-II to a gapped state, magnetoconductivity decreases, which is anticipated, consistent with previous experimental observations~\cite{wei2022ferromagnetic, zhang2017magnetic, 2025anomalies_inplane_hall}. This behavior is attributed to the Berry curvature, which modifies these observables during the topological transition from Weyl nodes to a gapped state. Our results provide a promising pathway for controlling topological Weyl crossings and their impact on the magnetic properties of the shandite systems.

In conclusion, we have shown that Co$_3$Sn$_2$S$_{(2-x)}$Se$_x$ can support spin-chiral interaction beyond the Dzyaloshinskii-Moriya interaction, due to the formation of a magnetic Kagome net, and this contribution remains a dominant energy term in the spin-orbit interaction. For the non-relativistic term, we calculate the exchange stiffness for all three compounds, and the results match well with the experimental data. By controlling the magnetic configurations, Weyl nodes in the shandite systems undergo a topological transition in the non-collinear case, whereas Weyl crossings remain unchanged in the non-coplanar case. The tunability of magnetic Weyl cones near the Fermi energy is confirmed by observing a topological phase transition from Type-I to Type-II, ultimately reaching a gapped state. Our study of magnetic interactions and Weyl-point manipulation via complex spin structures may open a novel approach to enhancing Berry curvature and the quantum-geometric term.

{\it Methods}-To investigate how magnetic textures affect the topological band structure of Co$_3$Sn$_2$S$_{(2-x)}$Se$_x$, as shown in Fig.\ref{FIG-1}.a, we use the full-potential linearized augmented plane wave (FLAPW) method in the  {\small FLEUR} code~\cite{fleurWeb, FLAPW}. We employ Perdew-Burke-Ernzerhof exchange functionals~\cite{perdew1996generalized} for approximating exchange-correlation energy. For all three materials, the plane-wave cutoff is $4.0 a.u^{-1}$. The k-point mesh is $26\times26\times26$ for all compounds. These criteria ensure self-consistent calculations for coplanar, noncoplanar, and magnetic anisotropy cases. 

Our shandite compounds are in bulk form and have a rhombohedral crystal structure. The muffin tin radii are 2.28 Bohr for Co, 2.75 Bohr for Sn, 1.78 Bohr for S, and 1.95 Bohr for Se. The lattice parameters are as follows Co$_3$Sn$_2$S$_2$ has $a=5.3747~\text{\AA}$ and $\alpha = 59.964 ^\circ$; Co$_3$Sn$_2$SeS has $a=5.47483~\text{\AA}$ and $\alpha = 59.13 ^\circ$; Co$_3$Sn$_2$Se$_2$ has $a=5.5635~\text{\AA}$ and $\alpha = 58.3^\circ$. Both Co$_3$Sn$_2$S$_2$ and Co$_3$Sn$_2$Se$_2$ are centrosymmetric at the Sn site $[0.5,0.5,0.5]$ in the Wyckoff position. In contrast, Co$_3$Sn$_2$SeS lacks this symmetry because Se and S atoms are substituted. All structural parameters come from experimental data~\cite{weihrich2015tuneable}. Collinear ferromagnetic calculations for all Co$_3$Sn$_2$S$_{(2-x)}$Se$_x$ compounds yield spin moment of $1.08 \mu_B$ and orbital moment of $-0.02 \mu_B$. Evaluating magnetic interactions requires self-consistent calculations to obtain energy curves for the homogeneous spin spiral and the noncoplanar scheme. To further probe the role of SOC in Co$_3$Sn$_2$S$_{(2-x)}$Se$_x$, we use first-order correction in noncoplanar case embedded in  {\small FLEUR} code ~\cite{fleurWeb,FLAPW}. For noncoplanar configurations, as in Fig.\ref{FIG-1}.b, the magnetic spin is described by changing the angle $\theta$ as follows:

\begin{equation}
\begin{aligned}
S_i=(\sin \phi_i \cos\theta_i; \sin \phi_i \sin \theta_i; \cos \theta_i)
\label{eq2}
\end{aligned}
\end{equation}

Starting from the ferromagnetic configuration, we consider three distinct spin moment directions, $\vec{S}_i$, for magnetic Co atoms. Here, $\theta_i = \{ \pi/2 - \theta; 0; 0 \}$  and $\phi_i= \{0; \theta/2;  -\theta/2\}$ are for azimuthal and polar angles for Co atoms, respectively. As a result, the spin-chiral interaction can be expressed as

\begin{equation}
\begin{aligned}
E_{SC} & = -\sum_i \kappa_i^{SC} \vec{L}^{TO}_i \cdot\vec{S}_i \\
& = -\sum_{i(jk)} \kappa^{SC}_{ijk}[\vec{S}_i(\vec{S}_j \times \vec{S}_k)](\vec{\tau}_{ijk} \cdot\vec{S}_i) \\
\label{eq3}
\end{aligned}
\end{equation} 

The magnetic structure defined in Eq.(\ref{eq2}) produces a nonzero scalar chiral product $\chi_{ijk} = \sin ^2 (\theta)$. This aligns the topological-orbital moment $\vec{L}^{TO}_i$ with the normal surface of the triangle spin lattice $\tau_{ijk}$ of the Kagome net. Since the normal surface vector of the kagome net points along the z-direction, the SOC projection is also taken along z-direction. The spin-chiral interaction strength is given by $\kappa^{SC}_{ijk}=\chi^{SO}_{ijk}\kappa^{TO}_{ijk}$, where the scalar quantities originate from the topological orbital moment $L^{TO}_i$ of each magnetic Co atom~\cite{grytsiuk2020topological}. The energy function described SCI as the function of $\theta$ can be written as $E_{SCI} = -\kappa^{SC}\sin^3{\theta}$.

To understand the behavior of $E_{MCA}$ in three-dimensional spheres, we calculate the energy difference between different angles of the magnetization axis in this system, which can be expressed by the following formula\cite{doring1957richtungsabhangigkeit}:

\begin{equation}
\begin{aligned}
E_{\text{total}} &= K_0 + K_1 \alpha_3^2 \\
&\quad + K_2 \left(\alpha_1^6 - 15 \alpha_1^4 \alpha_2^2 + 15 \alpha_1^2 \alpha_2^4 - \alpha_2^6 \right) \\
&\quad + K_3 \alpha_3 \left(\alpha_1^3 - 3 \alpha_1 \alpha_2^2 \right)
\end{aligned}
\end{equation}

where $\alpha_1, \alpha_2$ and $\alpha_3$ are the direction cosines of the spontaneous magnetization. Under the projection along z-direction, the magneticrystalline anisotropic energy can be written as $E_{MAE}=-K_1\left[1-\cos^2(\theta)\right]$, where $K_1$ is the dominant term in MAE as indicated in the main text. In addition, since only $Co_1$ has out-of-plane components $S_1=\left( 0,0,\cos\left( \frac{\pi}{2}-\theta \right)\right)$, we simplify the contribution of $K_1$ of $E_{MAE}$ is only one-third compared to the strength of MAE. Similarly, DM energy variation along z-direction can be described as $E_{DMI}=\sum_{ij}D_{ij}S_i \times S_j=-D\sin(\theta)$. Further details of the derivation of these energy functions can be found in S1 section from SI. The fitting energy functions for our shandite system under the effect of SOC in noncoplanar spin configurations can be written as follows:

\begin{equation}
\begin{aligned}
\Delta E_{SOC} &= E_{SCI}+E_{MAE}+E_{DMI} \\
&\quad =\kappa^{SC}\sin^3\theta - K_1 \sin^2\theta -D\sin\theta
\label{eq4}
\end{aligned}
\end{equation}


\section{Acknowledgements}
  This work at USF was based upon work supported by the Air Force Office of Scientific Research under award number FA9550-23-1-0132 and US Department of Energy (DOE) under Grant No. DE-FG0207ER46438. J.G. and D.K.L. acknowledges support from the Max Planck Society through the Max Planck Partner Group Programme and CS-CSA-2024-106 from Research Corporation for Science Advancement. Y.M. acknowledges support by the Deutsche Forschungsgemeinschaft (DFG) in the framework of TRR 288/2 $-$ 422213477 (Project B06) and by the J\"{u}lich Supercomputing Centre (jiff40).

%


\begin{thebibliography}{60}%
\makeatletter
\providecommand \@ifxundefined [1]{%
 \@ifx{#1\undefined}
}%
\providecommand \@ifnum [1]{%
 \ifnum #1\expandafter \@firstoftwo
 \else \expandafter \@secondoftwo
 \fi
}%
\providecommand \@ifx [1]{%
 \ifx #1\expandafter \@firstoftwo
 \else \expandafter \@secondoftwo
 \fi
}%
\providecommand \natexlab [1]{#1}%
\providecommand \enquote  [1]{``#1''}%
\providecommand \bibnamefont  [1]{#1}%
\providecommand \bibfnamefont [1]{#1}%
\providecommand \citenamefont [1]{#1}%
\providecommand \@href[1]{\@@startlink{#1}\@@href}%
\providecommand \@@href[1]{\endgroup#1\@@endlink}%
\providecommand \@sanitize@url [0]{\catcode `\\12\catcode `\$12\catcode
  `\&12\catcode `\#12\catcode `\^12\catcode `\_12\catcode `\%12\relax}%
\providecommand \@@startlink[1]{}%
\providecommand \@@endlink[0]{}%
\providecommand \@url [1]{\endgroup\@href {#1}{\urlprefix }}%
\providecommand \urlprefix  [0]{URL }%
\providecommand \doibase [0]{http://dx.doi.org/}%
\providecommand \selectlanguage [0]{\@gobble}%
\providecommand \bibinfo  [0]{\@secondoftwo}%
\providecommand \bibfield  [0]{\@secondoftwo}%
\providecommand \translation [1]{[#1]}%
\providecommand \BibitemOpen [0]{}%
\providecommand \bibitemStop [0]{}%
\providecommand \bibitemNoStop [0]{.\EOS\space}%
\providecommand \EOS [0]{\spacefactor3000\relax}%
\providecommand \BibitemShut  [1]{\csname bibitem#1\endcsname}%
\let\auto@bib@innerbib\@empty
\bibitem [{\citenamefont {Burkov}(2018)}]{BurkovRev2018}%
  \BibitemOpen
  \bibfield  {author} {\bibinfo {author} {\bibfnamefont {A.}~\bibnamefont
  {Burkov}},\ }{\bibfield  {journal} {\bibinfo  {journal} {Annual
  Review of Condensed Matter Physics}\ }\textbf {\bibinfo {volume} {9}},\
  \bibinfo {pages} {359} (\bibinfo {year} {2018})}\BibitemShut {NoStop}%
\bibitem [{\citenamefont {Bernevig}\ \emph {et~al.}(2022)\citenamefont
  {Bernevig}, \citenamefont {Felser},\ and\ \citenamefont
  {Beidenkopf}}]{Bernevig2022}%
  \BibitemOpen
  \bibfield  {author} {\bibinfo {author} {\bibfnamefont {B.~A.}\ \bibnamefont
  {Bernevig}}, \bibinfo {author} {\bibfnamefont {C.}~\bibnamefont {Felser}}, \
  and\ \bibinfo {author} {\bibfnamefont {H.}~\bibnamefont {Beidenkopf}},\
  }\href {\doibase 10.1038/s41586-021-04105-x} {\bibfield  {journal} {\bibinfo
  {journal} {Nature}\ }\textbf {\bibinfo {volume} {603}},\ \bibinfo {pages}
  {41} (\bibinfo {year} {2022})}\BibitemShut {NoStop}%
\bibitem [{\citenamefont {Yan}\ and\ \citenamefont
  {Felser}(2017)}]{TopmatYan2017}%
  \BibitemOpen
  \bibfield  {author} {\bibinfo {author} {\bibfnamefont {B.}~\bibnamefont
  {Yan}}\ and\ \bibinfo {author} {\bibfnamefont {C.}~\bibnamefont {Felser}},\
  }{\bibfield  {journal} {\bibinfo  {journal} {Annual Review of
  Condensed Matter Physics}\ }\textbf {\bibinfo {volume} {8}},\ \bibinfo
  {pages} {337} (\bibinfo {year} {2017})}\BibitemShut {NoStop}%
\bibitem [{\citenamefont {Ohta}\ \emph {et~al.}(2006)\citenamefont {Ohta},
  \citenamefont {Bostwick}, \citenamefont {Seyller}, \citenamefont {Horn},\
  and\ \citenamefont {Rotenberg}}]{Ohta2006}%
  \BibitemOpen
  \bibfield  {author} {\bibinfo {author} {\bibfnamefont {T.}~\bibnamefont
  {Ohta}}, \bibinfo {author} {\bibfnamefont {A.}~\bibnamefont {Bostwick}},
  \bibinfo {author} {\bibfnamefont {T.}~\bibnamefont {Seyller}}, \bibinfo
  {author} {\bibfnamefont {K.}~\bibnamefont {Horn}}, \ and\ \bibinfo {author}
  {\bibfnamefont {E.}~\bibnamefont {Rotenberg}},\ }{\bibfield
  {journal} {\bibinfo  {journal} {Science}\ }\textbf {\bibinfo {volume}
  {313}},\ \bibinfo {pages} {951} (\bibinfo {year} {2006})}\BibitemShut
  {NoStop}%
\bibitem [{\citenamefont {Cao}\ \emph {et~al.}(2018)\citenamefont {Cao},
  \citenamefont {Fatemi}, \citenamefont {Fang}, \citenamefont {Watanabe},
  \citenamefont {Taniguchi}, \citenamefont {Kaxiras},\ and\ \citenamefont
  {Jarillo-Herrero}}]{Cao2018}%
  \BibitemOpen
  \bibfield  {author} {\bibinfo {author} {\bibfnamefont {Y.}~\bibnamefont
  {Cao}}, \bibinfo {author} {\bibfnamefont {V.}~\bibnamefont {Fatemi}},
  \bibinfo {author} {\bibfnamefont {S.}~\bibnamefont {Fang}}, \bibinfo {author}
  {\bibfnamefont {K.}~\bibnamefont {Watanabe}}, \bibinfo {author}
  {\bibfnamefont {T.}~\bibnamefont {Taniguchi}}, \bibinfo {author}
  {\bibfnamefont {E.}~\bibnamefont {Kaxiras}}, \ and\ \bibinfo {author}
  {\bibfnamefont {P.}~\bibnamefont {Jarillo-Herrero}},\ }
  {\bibfield  {journal} {\bibinfo  {journal} {Nature}\ }\textbf {\bibinfo
  {volume} {556}},\ \bibinfo {pages} {43} (\bibinfo {year} {2018})}\BibitemShut
  {NoStop}%
\bibitem [{\citenamefont {Andrei}\ and\ \citenamefont
  {MacDonald}(2020)}]{Andrei2020}%
  \BibitemOpen
  \bibfield  {author} {\bibinfo {author} {\bibfnamefont {E.~Y.}\ \bibnamefont
  {Andrei}}\ and\ \bibinfo {author} {\bibfnamefont {A.~H.}\ \bibnamefont
  {MacDonald}},\ }\href {\doibase 10.1038/s41563-020-00840-0} {\bibfield
  {journal} {\bibinfo  {journal} {Nature Materials}\ }\textbf {\bibinfo
  {volume} {19}},\ \bibinfo {pages} {1265} (\bibinfo {year}
  {2020})}\BibitemShut {NoStop}%
\bibitem [{\citenamefont {Gooth}\ \emph {et~al.}(2019)\citenamefont {Gooth},
  \citenamefont {Bradlyn}, \citenamefont {Honnali}, \citenamefont {Schindler},
  \citenamefont {Kumar}, \citenamefont {Noky}, \citenamefont {Qi},
  \citenamefont {Shekhar}, \citenamefont {Sun}, \citenamefont {Wang},
  \citenamefont {Bernevig},\ and\ \citenamefont {Felser}}]{Gooth2019}%
  \BibitemOpen
  \bibfield  {author} {\bibinfo {author} {\bibfnamefont {J.}~\bibnamefont
  {Gooth}}, \bibinfo {author} {\bibfnamefont {B.}~\bibnamefont {Bradlyn}},
  \bibinfo {author} {\bibfnamefont {S.}~\bibnamefont {Honnali}}, \bibinfo
  {author} {\bibfnamefont {C.}~\bibnamefont {Schindler}}, \bibinfo {author}
  {\bibfnamefont {N.}~\bibnamefont {Kumar}}, \bibinfo {author} {\bibfnamefont
  {J.}~\bibnamefont {Noky}}, \bibinfo {author} {\bibfnamefont {Y.}~\bibnamefont
  {Qi}}, \bibinfo {author} {\bibfnamefont {C.}~\bibnamefont {Shekhar}},
  \bibinfo {author} {\bibfnamefont {Y.}~\bibnamefont {Sun}}, \bibinfo {author}
  {\bibfnamefont {Z.}~\bibnamefont {Wang}}, \bibinfo {author} {\bibfnamefont
  {B.~A.}\ \bibnamefont {Bernevig}}, \ and\ \bibinfo {author} {\bibfnamefont
  {C.}~\bibnamefont {Felser}},\ }{\bibfield  {journal} {\bibinfo
  {journal} {Nature}\ }\textbf {\bibinfo {volume} {575}},\ \bibinfo {pages}
  {315} (\bibinfo {year} {2019})}\BibitemShut {NoStop}%
\bibitem [{\citenamefont {Shi}\ \emph {et~al.}(2021)\citenamefont {Shi},
  \citenamefont {Wieder}, \citenamefont {Meyerheim}, \citenamefont {Sun},
  \citenamefont {Zhang}, \citenamefont {Li}, \citenamefont {Shen},
  \citenamefont {Qi}, \citenamefont {Yang}, \citenamefont {Jena}, \citenamefont
  {Werner}, \citenamefont {Koepernik}, \citenamefont {Parkin}, \citenamefont
  {Chen}, \citenamefont {Felser}, \citenamefont {Bernevig},\ and\ \citenamefont
  {Wang}}]{Shi2021}%
  \BibitemOpen
  \bibfield  {author} {\bibinfo {author} {\bibfnamefont {W.}~\bibnamefont
  {Shi}}, \bibinfo {author} {\bibfnamefont {B.~J.}\ \bibnamefont {Wieder}},
  \bibinfo {author} {\bibfnamefont {H.~L.}\ \bibnamefont {Meyerheim}}, \bibinfo
  {author} {\bibfnamefont {Y.}~\bibnamefont {Sun}}, \bibinfo {author}
  {\bibfnamefont {Y.}~\bibnamefont {Zhang}}, \bibinfo {author} {\bibfnamefont
  {Y.}~\bibnamefont {Li}}, \bibinfo {author} {\bibfnamefont {L.}~\bibnamefont
  {Shen}}, \bibinfo {author} {\bibfnamefont {Y.}~\bibnamefont {Qi}}, \bibinfo
  {author} {\bibfnamefont {L.}~\bibnamefont {Yang}}, \bibinfo {author}
  {\bibfnamefont {J.}~\bibnamefont {Jena}}, \bibinfo {author} {\bibfnamefont
  {P.}~\bibnamefont {Werner}}, \bibinfo {author} {\bibfnamefont
  {K.}~\bibnamefont {Koepernik}}, \bibinfo {author} {\bibfnamefont
  {S.}~\bibnamefont {Parkin}}, \bibinfo {author} {\bibfnamefont
  {Y.}~\bibnamefont {Chen}}, \bibinfo {author} {\bibfnamefont {C.}~\bibnamefont
  {Felser}}, \bibinfo {author} {\bibfnamefont {B.~A.}\ \bibnamefont
  {Bernevig}}, \ and\ \bibinfo {author} {\bibfnamefont {Z.}~\bibnamefont
  {Wang}},\ }{\bibfield  {journal} {\bibinfo  {journal} {Nature
  Physics}\ }\textbf {\bibinfo {volume} {17}},\ \bibinfo {pages} {381}
  (\bibinfo {year} {2021})}\BibitemShut {NoStop}%
\bibitem [{\citenamefont {Hanke}\ \emph {et~al.}(2017)\citenamefont {Hanke},
  \citenamefont {Freimuth}, \citenamefont {Niu}, \citenamefont {Bl{\"u}gel},\
  and\ \citenamefont {Mokrousov}}]{hanke2017mixed}%
  \BibitemOpen
  \bibfield  {author} {\bibinfo {author} {\bibfnamefont {J.-P.}\ \bibnamefont
  {Hanke}}, \bibinfo {author} {\bibfnamefont {F.}~\bibnamefont {Freimuth}},
  \bibinfo {author} {\bibfnamefont {C.}~\bibnamefont {Niu}}, \bibinfo {author}
  {\bibfnamefont {S.}~\bibnamefont {Bl{\"u}gel}}, \ and\ \bibinfo {author}
  {\bibfnamefont {Y.}~\bibnamefont {Mokrousov}},\ }{\bibfield
  {journal} {\bibinfo  {journal} {Nature Communications}\ }\textbf {\bibinfo
  {volume} {8}},\ \bibinfo {pages} {1479} (\bibinfo {year} {2017})}\BibitemShut
  {NoStop}%
\bibitem [{\citenamefont {Niu}\ \emph {et~al.}(2019)\citenamefont {Niu},
  \citenamefont {Hanke}, \citenamefont {Buhl}, \citenamefont {Zhang},
  \citenamefont {Plucinski}, \citenamefont {Wortmann}, \citenamefont
  {Bl{\"u}gel}, \citenamefont {Bihlmayer},\ and\ \citenamefont
  {Mokrousov}}]{Niu2019}%
  \BibitemOpen
  \bibfield  {author} {\bibinfo {author} {\bibfnamefont {C.}~\bibnamefont
  {Niu}}, \bibinfo {author} {\bibfnamefont {J.-P.}\ \bibnamefont {Hanke}},
  \bibinfo {author} {\bibfnamefont {P.~M.}\ \bibnamefont {Buhl}}, \bibinfo
  {author} {\bibfnamefont {H.}~\bibnamefont {Zhang}}, \bibinfo {author}
  {\bibfnamefont {L.}~\bibnamefont {Plucinski}}, \bibinfo {author}
  {\bibfnamefont {D.}~\bibnamefont {Wortmann}}, \bibinfo {author}
  {\bibfnamefont {S.}~\bibnamefont {Bl{\"u}gel}}, \bibinfo {author}
  {\bibfnamefont {G.}~\bibnamefont {Bihlmayer}}, \ and\ \bibinfo {author}
  {\bibfnamefont {Y.}~\bibnamefont {Mokrousov}},\ }{\bibfield
  {journal} {\bibinfo  {journal} {Nature Communications}\ }\textbf {\bibinfo
  {volume} {10}},\ \bibinfo {pages} {3179} (\bibinfo {year}
  {2019})}\BibitemShut {NoStop}%
\bibitem [{\citenamefont {Dzyaloshinsky}(1958)}]{Dzyaloshinsky1958}%
  \BibitemOpen
  \bibfield  {author} {\bibinfo {author} {\bibfnamefont {I.}~\bibnamefont
  {Dzyaloshinsky}},\ }{\bibfield  {journal} {\bibinfo  {journal}
  {Journal of Physics and Chemistry of Solids}\ }\textbf {\bibinfo {volume}
  {4}},\ \bibinfo {pages} {241} (\bibinfo {year} {1958})}\BibitemShut {NoStop}%
\bibitem [{\citenamefont {Moriya}(1960)}]{Moriya1960}%
  \BibitemOpen
  \bibfield  {author} {\bibinfo {author} {\bibfnamefont {T.}~\bibnamefont
  {Moriya}},\ }{\bibfield  {journal} {\bibinfo  {journal} {Phys.
  Rev.}\ }\textbf {\bibinfo {volume} {120}},\ \bibinfo {pages} {91} (\bibinfo
  {year} {1960})}\BibitemShut {NoStop}%
\bibitem [{\citenamefont {Freimuth}\ \emph {et~al.}(2013)\citenamefont
  {Freimuth}, \citenamefont {Bamler}, \citenamefont {Mokrousov},\ and\
  \citenamefont {Rosch}}]{freimuth2013phase}%
  \BibitemOpen
  \bibfield  {author} {\bibinfo {author} {\bibfnamefont {F.}~\bibnamefont
  {Freimuth}}, \bibinfo {author} {\bibfnamefont {R.}~\bibnamefont {Bamler}},
  \bibinfo {author} {\bibfnamefont {Y.}~\bibnamefont {Mokrousov}}, \ and\
  \bibinfo {author} {\bibfnamefont {A.}~\bibnamefont {Rosch}},\ }
  {\bibfield  {journal} {\bibinfo  {journal} {arXiv preprint arXiv:1307.8085}\
  } (\bibinfo {year} {2013})}\BibitemShut {NoStop}%
\bibitem [{\citenamefont {Bogdanov}\ and\ \citenamefont
  {Yablonskii}(1989)}]{bogdanov1989}%
  \BibitemOpen
  \bibfield  {author} {\bibinfo {author} {\bibfnamefont {A.~N.}\ \bibnamefont
  {Bogdanov}}\ and\ \bibinfo {author} {\bibfnamefont {D.}~\bibnamefont
  {Yablonskii}},\ }{\bibfield  {journal} {\bibinfo  {journal}
  {Zh. Eksp. Teor. Fiz}\ }\textbf {\bibinfo {volume} {95}},\ \bibinfo {pages}
  {178} (\bibinfo {year} {1989})}\BibitemShut {NoStop}%
\bibitem [{\citenamefont {M{\"u}hlbauer}\ \emph {et~al.}(2009)\citenamefont
  {M{\"u}hlbauer}, \citenamefont {Binz}, \citenamefont {Jonietz}, \citenamefont
  {Pfleiderer}, \citenamefont {Rosch}, \citenamefont {Neubauer}, \citenamefont
  {Georgii},\ and\ \citenamefont {B{\"o}ni}}]{Pfleiderer2009}%
  \BibitemOpen
  \bibfield  {author} {\bibinfo {author} {\bibfnamefont {S.}~\bibnamefont
  {M{\"u}hlbauer}}, \bibinfo {author} {\bibfnamefont {B.}~\bibnamefont {Binz}},
  \bibinfo {author} {\bibfnamefont {F.}~\bibnamefont {Jonietz}}, \bibinfo
  {author} {\bibfnamefont {C.}~\bibnamefont {Pfleiderer}}, \bibinfo {author}
  {\bibfnamefont {A.}~\bibnamefont {Rosch}}, \bibinfo {author} {\bibfnamefont
  {A.}~\bibnamefont {Neubauer}}, \bibinfo {author} {\bibfnamefont
  {R.}~\bibnamefont {Georgii}}, \ and\ \bibinfo {author} {\bibfnamefont
  {P.}~\bibnamefont {B{\"o}ni}},\ }{\bibfield  {journal}
  {\bibinfo  {journal} {Science}\ }\textbf {\bibinfo {volume} {323}},\ \bibinfo
  {pages} {915} (\bibinfo {year} {2009})}\BibitemShut {NoStop}%
\bibitem [{\citenamefont {Parkin}\ \emph {et~al.}(2008)\citenamefont {Parkin},
  \citenamefont {Hayashi},\ and\ \citenamefont {Thomas}}]{Parkin2008}%
  \BibitemOpen
  \bibfield  {author} {\bibinfo {author} {\bibfnamefont {S.~S.~P.}\
  \bibnamefont {Parkin}}, \bibinfo {author} {\bibfnamefont {M.}~\bibnamefont
  {Hayashi}}, \ and\ \bibinfo {author} {\bibfnamefont {L.}~\bibnamefont
  {Thomas}},\ }{\bibfield  {journal} {\bibinfo  {journal}
  {Science}\ }\textbf {\bibinfo {volume} {320}},\ \bibinfo {pages} {190}
  (\bibinfo {year} {2008})}\BibitemShut {NoStop}%
\bibitem [{\citenamefont {Tomasello}\ \emph {et~al.}(2014)\citenamefont
  {Tomasello}, \citenamefont {Martinez}, \citenamefont {Zivieri}, \citenamefont
  {Torres}, \citenamefont {Carpentieri},\ and\ \citenamefont
  {Finocchio}}]{Tomasello2014}%
  \BibitemOpen
  \bibfield  {author} {\bibinfo {author} {\bibfnamefont {R.}~\bibnamefont
  {Tomasello}}, \bibinfo {author} {\bibfnamefont {E.}~\bibnamefont {Martinez}},
  \bibinfo {author} {\bibfnamefont {R.}~\bibnamefont {Zivieri}}, \bibinfo
  {author} {\bibfnamefont {L.}~\bibnamefont {Torres}}, \bibinfo {author}
  {\bibfnamefont {M.}~\bibnamefont {Carpentieri}}, \ and\ \bibinfo {author}
  {\bibfnamefont {G.}~\bibnamefont {Finocchio}},\ }{\bibfield
  {journal} {\bibinfo  {journal} {Scientific Reports}\ }\textbf {\bibinfo
  {volume} {4}},\ \bibinfo {pages} {6784} (\bibinfo {year} {2014})}\BibitemShut
  {NoStop}%
\bibitem [{\citenamefont {Bruno}\ \emph {et~al.}(2004)\citenamefont {Bruno},
  \citenamefont {Dugaev},\ and\ \citenamefont {Taillefumier}}]{Bruno2004}%
  \BibitemOpen
  \bibfield  {author} {\bibinfo {author} {\bibfnamefont {P.}~\bibnamefont
  {Bruno}}, \bibinfo {author} {\bibfnamefont {V.~K.}\ \bibnamefont {Dugaev}}, \
  and\ \bibinfo {author} {\bibfnamefont {M.}~\bibnamefont {Taillefumier}},\
  }{\bibfield  {journal} {\bibinfo  {journal} {Phys. Rev. Lett.}\
  }\textbf {\bibinfo {volume} {93}},\ \bibinfo {pages} {096806} (\bibinfo
  {year} {2004})}\BibitemShut {NoStop}%
\bibitem [{\citenamefont {Kimbell}\ \emph {et~al.}(2022)\citenamefont
  {Kimbell}, \citenamefont {Kim}, \citenamefont {Wu}, \citenamefont {Cuoco},\
  and\ \citenamefont {Robinson}}]{Kimbell2022}%
  \BibitemOpen
  \bibfield  {author} {\bibinfo {author} {\bibfnamefont {G.}~\bibnamefont
  {Kimbell}}, \bibinfo {author} {\bibfnamefont {C.}~\bibnamefont {Kim}},
  \bibinfo {author} {\bibfnamefont {W.}~\bibnamefont {Wu}}, \bibinfo {author}
  {\bibfnamefont {M.}~\bibnamefont {Cuoco}}, \ and\ \bibinfo {author}
  {\bibfnamefont {J.~W.~A.}\ \bibnamefont {Robinson}},\ }
  {\bibfield  {journal} {\bibinfo  {journal} {Communications Materials}\
  }\textbf {\bibinfo {volume} {3}},\ \bibinfo {pages} {19} (\bibinfo {year}
  {2022})}\BibitemShut {NoStop}%
\bibitem [{\citenamefont {Heide}\ \emph {et~al.}(2008)\citenamefont {Heide},
  \citenamefont {Bihlmayer},\ and\ \citenamefont
  {Bl{\"u}gel}}]{heide2008dzyaloshinskii}%
  \BibitemOpen
  \bibfield  {author} {\bibinfo {author} {\bibfnamefont {M.}~\bibnamefont
  {Heide}}, \bibinfo {author} {\bibfnamefont {G.}~\bibnamefont {Bihlmayer}}, \
  and\ \bibinfo {author} {\bibfnamefont {S.}~\bibnamefont {Bl{\"u}gel}},\
  }{\bibfield  {journal} {\bibinfo  {journal} {Physical Review
  B---Condensed Matter and Materials Physics}\ }\textbf {\bibinfo {volume}
  {78}},\ \bibinfo {pages} {140403} (\bibinfo {year} {2008})}\BibitemShut
  {NoStop}%
\bibitem [{\citenamefont {Wang}\ \emph {et~al.}(2018)\citenamefont {Wang},
  \citenamefont {Yuan},\ and\ \citenamefont {Wang}}]{wang2018theory}%
  \BibitemOpen
  \bibfield  {author} {\bibinfo {author} {\bibfnamefont {X.}~\bibnamefont
  {Wang}}, \bibinfo {author} {\bibfnamefont {H.}~\bibnamefont {Yuan}}, \ and\
  \bibinfo {author} {\bibfnamefont {X.}~\bibnamefont {Wang}},\ }
  {\bibfield  {journal} {\bibinfo  {journal} {Communications Physics}\ }\textbf
  {\bibinfo {volume} {1}},\ \bibinfo {pages} {31} (\bibinfo {year}
  {2018})}\BibitemShut {NoStop}%
\bibitem [{\citenamefont {Jiang}\ \emph {et~al.}(2025)\citenamefont {Jiang},
  \citenamefont {Zhang},\ and\ \citenamefont {Zhou}}]{jiang2025higher}%
  \BibitemOpen
  \bibfield  {author} {\bibinfo {author} {\bibfnamefont {H.}~\bibnamefont
  {Jiang}}, \bibinfo {author} {\bibfnamefont {J.}~\bibnamefont {Zhang}}, \ and\
  \bibinfo {author} {\bibfnamefont {Y.}~\bibnamefont {Zhou}},\ }
  {\bibfield  {journal} {\bibinfo  {journal} {Physical Review B}\ }\textbf
  {\bibinfo {volume} {111}},\ \bibinfo {pages} {214429} (\bibinfo {year}
  {2025})}\BibitemShut {NoStop}%
\bibitem [{\citenamefont {Grytsiuk}\ \emph {et~al.}(2020)\citenamefont
  {Grytsiuk}, \citenamefont {Hanke}, \citenamefont {Hoffmann}, \citenamefont
  {Bouaziz}, \citenamefont {Gomonay}, \citenamefont {Bihlmayer}, \citenamefont
  {Lounis}, \citenamefont {Mokrousov},\ and\ \citenamefont
  {Bl{\"u}gel}}]{grytsiuk2020topological}%
  \BibitemOpen
  \bibfield  {author} {\bibinfo {author} {\bibfnamefont {S.}~\bibnamefont
  {Grytsiuk}}, \bibinfo {author} {\bibfnamefont {J.-P.}\ \bibnamefont {Hanke}},
  \bibinfo {author} {\bibfnamefont {M.}~\bibnamefont {Hoffmann}}, \bibinfo
  {author} {\bibfnamefont {J.}~\bibnamefont {Bouaziz}}, \bibinfo {author}
  {\bibfnamefont {O.}~\bibnamefont {Gomonay}}, \bibinfo {author} {\bibfnamefont
  {G.}~\bibnamefont {Bihlmayer}}, \bibinfo {author} {\bibfnamefont
  {S.}~\bibnamefont {Lounis}}, \bibinfo {author} {\bibfnamefont
  {Y.}~\bibnamefont {Mokrousov}}, \ and\ \bibinfo {author} {\bibfnamefont
  {S.}~\bibnamefont {Bl{\"u}gel}},\ }{\bibfield  {journal}
  {\bibinfo  {journal} {Nature communications}\ }\textbf {\bibinfo {volume}
  {11}},\ \bibinfo {pages} {511} (\bibinfo {year} {2020})}\BibitemShut
  {NoStop}%
\bibitem [{\citenamefont {Singh}\ \emph {et~al.}(2024)\citenamefont {Singh},
  \citenamefont {Jamaluddin}, \citenamefont {Pradhan}, \citenamefont {Nandy},
  \citenamefont {Tokunaga}, \citenamefont {Avdeev},\ and\ \citenamefont
  {Nayak}}]{singh2024higher}%
  \BibitemOpen
  \bibfield  {author} {\bibinfo {author} {\bibfnamefont {C.}~\bibnamefont
  {Singh}}, \bibinfo {author} {\bibfnamefont {S.}~\bibnamefont {Jamaluddin}},
  \bibinfo {author} {\bibfnamefont {S.}~\bibnamefont {Pradhan}}, \bibinfo
  {author} {\bibfnamefont {A.~K.}\ \bibnamefont {Nandy}}, \bibinfo {author}
  {\bibfnamefont {M.}~\bibnamefont {Tokunaga}}, \bibinfo {author}
  {\bibfnamefont {M.}~\bibnamefont {Avdeev}}, \ and\ \bibinfo {author}
  {\bibfnamefont {A.~K.}\ \bibnamefont {Nayak}},\ }{\bibfield
  {journal} {\bibinfo  {journal} {npj Quantum Materials}\ }\textbf {\bibinfo
  {volume} {9}},\ \bibinfo {pages} {43} (\bibinfo {year} {2024})}\BibitemShut
  {NoStop}%
\bibitem [{\citenamefont {Kolincio}\ \emph {et~al.}(2023)\citenamefont
  {Kolincio}, \citenamefont {Hirschberger}, \citenamefont {Masell},
  \citenamefont {Arima}, \citenamefont {Nagaosa},\ and\ \citenamefont
  {Tokura}}]{kolincio2023kagome}%
  \BibitemOpen
  \bibfield  {author} {\bibinfo {author} {\bibfnamefont {K.~K.}\ \bibnamefont
  {Kolincio}}, \bibinfo {author} {\bibfnamefont {M.}~\bibnamefont
  {Hirschberger}}, \bibinfo {author} {\bibfnamefont {J.}~\bibnamefont
  {Masell}}, \bibinfo {author} {\bibfnamefont {T.-h.}\ \bibnamefont {Arima}},
  \bibinfo {author} {\bibfnamefont {N.}~\bibnamefont {Nagaosa}}, \ and\
  \bibinfo {author} {\bibfnamefont {Y.}~\bibnamefont {Tokura}},\ }
  {\bibfield  {journal} {\bibinfo  {journal} {Physical Review Letters}\
  }\textbf {\bibinfo {volume} {130}},\ \bibinfo {pages} {136701} (\bibinfo
  {year} {2023})}\BibitemShut {NoStop}%
\bibitem [{\citenamefont {Zhang}\ \emph {et~al.}(2025)\citenamefont {Zhang},
  \citenamefont {Zhou},\ and\ \citenamefont {Cheng}}]{zhang2025valley}%
  \BibitemOpen
  \bibfield  {author} {\bibinfo {author} {\bibfnamefont {W.}~\bibnamefont
  {Zhang}}, \bibinfo {author} {\bibfnamefont {J.}~\bibnamefont {Zhou}}, \ and\
  \bibinfo {author} {\bibfnamefont {S.}~\bibnamefont {Cheng}},\ }
  {\bibfield  {journal} {\bibinfo  {journal} {Physical Review B}\ }\textbf
  {\bibinfo {volume} {111}},\ \bibinfo {pages} {165307} (\bibinfo {year}
  {2025})}\BibitemShut {NoStop}%
\bibitem [{\citenamefont {Yu}\ \emph {et~al.}(2025)\citenamefont {Yu},
  \citenamefont {Bernevig}, \citenamefont {Queiroz}, \citenamefont {Rossi},
  \citenamefont {T{\"o}rm{\"a}},\ and\ \citenamefont {Yang}}]{yu2025quantum}%
  \BibitemOpen
  \bibfield  {author} {\bibinfo {author} {\bibfnamefont {J.}~\bibnamefont
  {Yu}}, \bibinfo {author} {\bibfnamefont {B.~A.}\ \bibnamefont {Bernevig}},
  \bibinfo {author} {\bibfnamefont {R.}~\bibnamefont {Queiroz}}, \bibinfo
  {author} {\bibfnamefont {E.}~\bibnamefont {Rossi}}, \bibinfo {author}
  {\bibfnamefont {P.}~\bibnamefont {T{\"o}rm{\"a}}}, \ and\ \bibinfo {author}
  {\bibfnamefont {B.-J.}\ \bibnamefont {Yang}},\ }{\bibfield
  {journal} {\bibinfo  {journal} {npj Quantum Materials}\ }\textbf {\bibinfo
  {volume} {10}},\ \bibinfo {pages} {101} (\bibinfo {year} {2025})}\BibitemShut
  {NoStop}%
\bibitem [{\citenamefont {van Walsem}\ \emph {et~al.}(2020)\citenamefont {van
  Walsem}, \citenamefont {Duine},\ and\ \citenamefont
  {Guimar{\~a}es}}]{van2020layer}%
  \BibitemOpen
  \bibfield  {author} {\bibinfo {author} {\bibfnamefont {E.}~\bibnamefont {van
  Walsem}}, \bibinfo {author} {\bibfnamefont {R.~A.}\ \bibnamefont {Duine}}, \
  and\ \bibinfo {author} {\bibfnamefont {M.~H.}\ \bibnamefont
  {Guimar{\~a}es}},\ }{\bibfield  {journal} {\bibinfo  {journal}
  {Physical Review B}\ }\textbf {\bibinfo {volume} {102}},\ \bibinfo {pages}
  {174403} (\bibinfo {year} {2020})}\BibitemShut {NoStop}%
\bibitem [{\citenamefont {Liu}\ \emph {et~al.}(2018)\citenamefont {Liu},
  \citenamefont {Sun}, \citenamefont {Kumar}, \citenamefont {Muechler},
  \citenamefont {Sun}, \citenamefont {Jiao}, \citenamefont {Yang},
  \citenamefont {Liu}, \citenamefont {Liang}, \citenamefont {Xu} \emph
  {et~al.}}]{liu2018giant}%
  \BibitemOpen
  \bibfield  {author} {\bibinfo {author} {\bibfnamefont {E.}~\bibnamefont
  {Liu}}, \bibinfo {author} {\bibfnamefont {Y.}~\bibnamefont {Sun}}, \bibinfo
  {author} {\bibfnamefont {N.}~\bibnamefont {Kumar}}, \bibinfo {author}
  {\bibfnamefont {L.}~\bibnamefont {Muechler}}, \bibinfo {author}
  {\bibfnamefont {A.}~\bibnamefont {Sun}}, \bibinfo {author} {\bibfnamefont
  {L.}~\bibnamefont {Jiao}}, \bibinfo {author} {\bibfnamefont {S.-Y.}\
  \bibnamefont {Yang}}, \bibinfo {author} {\bibfnamefont {D.}~\bibnamefont
  {Liu}}, \bibinfo {author} {\bibfnamefont {A.}~\bibnamefont {Liang}}, \bibinfo
  {author} {\bibfnamefont {Q.}~\bibnamefont {Xu}},  \emph {et~al.},\
  }{\bibfield  {journal} {\bibinfo  {journal} {Nature physics}\
  }\textbf {\bibinfo {volume} {14}},\ \bibinfo {pages} {1125} (\bibinfo {year}
  {2018})}\BibitemShut {NoStop}%
\bibitem [{\citenamefont {Mitscherling}(2020)}]{mitscherling2020longitudinal}%
  \BibitemOpen
  \bibfield  {author} {\bibinfo {author} {\bibfnamefont {J.}~\bibnamefont
  {Mitscherling}},\ }{\bibfield  {journal} {\bibinfo  {journal}
  {Physical Review B}\ }\textbf {\bibinfo {volume} {102}},\ \bibinfo {pages}
  {165151} (\bibinfo {year} {2020})}\BibitemShut {NoStop}%
\bibitem [{\citenamefont {Sakai}\ \emph {et~al.}(2013)\citenamefont {Sakai},
  \citenamefont {Kamihara},\ and\ \citenamefont {Matoba}}]{sakai2013magnetic}%
  \BibitemOpen
  \bibfield  {author} {\bibinfo {author} {\bibfnamefont {Y.}~\bibnamefont
  {Sakai}}, \bibinfo {author} {\bibfnamefont {Y.}~\bibnamefont {Kamihara}}, \
  and\ \bibinfo {author} {\bibfnamefont {M.}~\bibnamefont {Matoba}},\
  }{\bibfield  {journal} {\bibinfo  {journal} {physica status
  solidi c}\ }\textbf {\bibinfo {volume} {10}},\ \bibinfo {pages} {1130}
  (\bibinfo {year} {2013})}\BibitemShut {NoStop}%
\bibitem [{\citenamefont {Shin}\ \emph {et~al.}(2021)\citenamefont {Shin},
  \citenamefont {Jun}, \citenamefont {Lee},\ and\ \citenamefont
  {Jung}}]{shin2021Tc}%
  \BibitemOpen
  \bibfield  {author} {\bibinfo {author} {\bibfnamefont {D.-H.}\ \bibnamefont
  {Shin}}, \bibinfo {author} {\bibfnamefont {J.-H.}\ \bibnamefont {Jun}},
  \bibinfo {author} {\bibfnamefont {S.-E.}\ \bibnamefont {Lee}}, \ and\
  \bibinfo {author} {\bibfnamefont {M.-H.}\ \bibnamefont {Jung}},\ } 
  {\bibfield  {journal} {\bibinfo  {journal} {arXiv preprint
  arXiv:2105.03892}\ } (\bibinfo {year} {2021})}\BibitemShut {NoStop}%
\bibitem [{\citenamefont {Barua}\ \emph {et~al.}(2025)\citenamefont {Barua},
  \citenamefont {Liu}, \citenamefont {Knapp}, \citenamefont {Boegel},
  \citenamefont {Carrera}, \citenamefont {Mapara}, \citenamefont {Karaiskaj},
  \citenamefont {Romestan}, \citenamefont {Bhat}, \citenamefont {Romero} \emph
  {et~al.}}]{barua2025competing}%
  \BibitemOpen
  \bibfield  {author} {\bibinfo {author} {\bibfnamefont {A.}~\bibnamefont
  {Barua}}, \bibinfo {author} {\bibfnamefont {H.}~\bibnamefont {Liu}}, \bibinfo
  {author} {\bibfnamefont {S.}~\bibnamefont {Knapp}}, \bibinfo {author}
  {\bibfnamefont {C.}~\bibnamefont {Boegel}}, \bibinfo {author} {\bibfnamefont
  {S.~L.}\ \bibnamefont {Carrera}}, \bibinfo {author} {\bibfnamefont
  {V.}~\bibnamefont {Mapara}}, \bibinfo {author} {\bibfnamefont
  {D.}~\bibnamefont {Karaiskaj}}, \bibinfo {author} {\bibfnamefont
  {Z.}~\bibnamefont {Romestan}}, \bibinfo {author} {\bibfnamefont {S.~S.}\
  \bibnamefont {Bhat}}, \bibinfo {author} {\bibfnamefont {A.~H.}\ \bibnamefont
  {Romero}},  \emph {et~al.},\ }{\bibfield  {journal} {\bibinfo
  {journal} {physica status solidi (b)}\ ,\ \bibinfo {pages} {2400520}}
  (\bibinfo {year} {2025})}\BibitemShut {NoStop}%
\bibitem [{\citenamefont {Guguchia}\ \emph {et~al.}(2020)\citenamefont
  {Guguchia}, \citenamefont {Verezhak}, \citenamefont {Gawryluk}, \citenamefont
  {Tsirkin}, \citenamefont {Yin}, \citenamefont {Belopolski}, \citenamefont
  {Zhou}, \citenamefont {Simutis}, \citenamefont {Zhang}, \citenamefont
  {Cochran} \emph {et~al.}}]{guguchia2020tunable}%
  \BibitemOpen
  \bibfield  {author} {\bibinfo {author} {\bibfnamefont {Z.}~\bibnamefont
  {Guguchia}}, \bibinfo {author} {\bibfnamefont {J.}~\bibnamefont {Verezhak}},
  \bibinfo {author} {\bibfnamefont {D.}~\bibnamefont {Gawryluk}}, \bibinfo
  {author} {\bibfnamefont {S.}~\bibnamefont {Tsirkin}}, \bibinfo {author}
  {\bibfnamefont {J.-X.}\ \bibnamefont {Yin}}, \bibinfo {author} {\bibfnamefont
  {I.}~\bibnamefont {Belopolski}}, \bibinfo {author} {\bibfnamefont
  {H.}~\bibnamefont {Zhou}}, \bibinfo {author} {\bibfnamefont {G.}~\bibnamefont
  {Simutis}}, \bibinfo {author} {\bibfnamefont {S.-S.}\ \bibnamefont {Zhang}},
  \bibinfo {author} {\bibfnamefont {T.}~\bibnamefont {Cochran}},  \emph
  {et~al.},\ }{\bibfield  {journal} {\bibinfo  {journal} {Nature
  communications}\ }\textbf {\bibinfo {volume} {11}},\ \bibinfo {pages} {559}
  (\bibinfo {year} {2020})}\BibitemShut {NoStop}%
\bibitem [{\citenamefont {Wu}\ \emph {et~al.}(2020)\citenamefont {Wu},
  \citenamefont {Sun}, \citenamefont {Hsieh}, \citenamefont {Chen},
  \citenamefont {Kakarla}, \citenamefont {Deng}, \citenamefont {Chu},\ and\
  \citenamefont {Yang}}]{wu2020observation}%
  \BibitemOpen
  \bibfield  {author} {\bibinfo {author} {\bibfnamefont {H.}~\bibnamefont
  {Wu}}, \bibinfo {author} {\bibfnamefont {P.}~\bibnamefont {Sun}}, \bibinfo
  {author} {\bibfnamefont {D.}~\bibnamefont {Hsieh}}, \bibinfo {author}
  {\bibfnamefont {H.}~\bibnamefont {Chen}}, \bibinfo {author} {\bibfnamefont
  {D.~C.}\ \bibnamefont {Kakarla}}, \bibinfo {author} {\bibfnamefont
  {L.}~\bibnamefont {Deng}}, \bibinfo {author} {\bibfnamefont {C.}~\bibnamefont
  {Chu}}, \ and\ \bibinfo {author} {\bibfnamefont {H.}~\bibnamefont {Yang}},\
  }{\bibfield  {journal} {\bibinfo  {journal} {Materials Today
  Physics}\ }\textbf {\bibinfo {volume} {12}},\ \bibinfo {pages} {100189}
  (\bibinfo {year} {2020})}\BibitemShut {NoStop}%
\bibitem [{\citenamefont {Kassem}\ \emph {et~al.}(2017)\citenamefont {Kassem},
  \citenamefont {Tabata}, \citenamefont {Waki},\ and\ \citenamefont
  {Nakamura}}]{kassem2017low}%
  \BibitemOpen
  \bibfield  {author} {\bibinfo {author} {\bibfnamefont {M.~A.}\ \bibnamefont
  {Kassem}}, \bibinfo {author} {\bibfnamefont {Y.}~\bibnamefont {Tabata}},
  \bibinfo {author} {\bibfnamefont {T.}~\bibnamefont {Waki}}, \ and\ \bibinfo
  {author} {\bibfnamefont {H.}~\bibnamefont {Nakamura}},\ }
  {\bibfield  {journal} {\bibinfo  {journal} {Physical Review B}\ }\textbf
  {\bibinfo {volume} {96}},\ \bibinfo {pages} {014429} (\bibinfo {year}
  {2017})}\BibitemShut {NoStop}%
\bibitem [{\citenamefont {Fujiwara}\ \emph {et~al.}(2024)\citenamefont
  {Fujiwara}, \citenamefont {Ogawa}, \citenamefont {Yoshikawa}, \citenamefont
  {Kobayashi}, \citenamefont {Nomura}, \citenamefont {Shimano},\ and\
  \citenamefont {Tsukazaki}}]{fujiwara2024giant}%
  \BibitemOpen
  \bibfield  {author} {\bibinfo {author} {\bibfnamefont {K.}~\bibnamefont
  {Fujiwara}}, \bibinfo {author} {\bibfnamefont {K.}~\bibnamefont {Ogawa}},
  \bibinfo {author} {\bibfnamefont {N.}~\bibnamefont {Yoshikawa}}, \bibinfo
  {author} {\bibfnamefont {K.}~\bibnamefont {Kobayashi}}, \bibinfo {author}
  {\bibfnamefont {K.}~\bibnamefont {Nomura}}, \bibinfo {author} {\bibfnamefont
  {R.}~\bibnamefont {Shimano}}, \ and\ \bibinfo {author} {\bibfnamefont
  {A.}~\bibnamefont {Tsukazaki}},\ }{\bibfield  {journal}
  {\bibinfo  {journal} {Communications Materials}\ }\textbf {\bibinfo {volume}
  {5}},\ \bibinfo {pages} {239} (\bibinfo {year} {2024})}\BibitemShut {NoStop}%
\bibitem [{\citenamefont {Sugawara}\ \emph {et~al.}(2019)\citenamefont
  {Sugawara}, \citenamefont {Akashi}, \citenamefont {Kassem}, \citenamefont
  {Tabata}, \citenamefont {Waki},\ and\ \citenamefont
  {Nakamura}}]{sugawara2019magnetic}%
  \BibitemOpen
  \bibfield  {author} {\bibinfo {author} {\bibfnamefont {A.}~\bibnamefont
  {Sugawara}}, \bibinfo {author} {\bibfnamefont {T.}~\bibnamefont {Akashi}},
  \bibinfo {author} {\bibfnamefont {M.~A.}\ \bibnamefont {Kassem}}, \bibinfo
  {author} {\bibfnamefont {Y.}~\bibnamefont {Tabata}}, \bibinfo {author}
  {\bibfnamefont {T.}~\bibnamefont {Waki}}, \ and\ \bibinfo {author}
  {\bibfnamefont {H.}~\bibnamefont {Nakamura}},\ }{\bibfield
  {journal} {\bibinfo  {journal} {Physical Review Materials}\ }\textbf
  {\bibinfo {volume} {3}},\ \bibinfo {pages} {104421} (\bibinfo {year}
  {2019})}\BibitemShut {NoStop}%
\bibitem [{\citenamefont {Pate}\ \emph
  {et~al.}(2025{\natexlab{a}})\citenamefont {Pate}, \citenamefont {Wang},
  \citenamefont {Shen}, \citenamefont {Xiao}, \citenamefont {Welp},\ and\
  \citenamefont {Vlasko-Vlasov}}]{pate2025anomalies}%
  \BibitemOpen
  \bibfield  {author} {\bibinfo {author} {\bibfnamefont {S.}~\bibnamefont
  {Pate}}, \bibinfo {author} {\bibfnamefont {B.}~\bibnamefont {Wang}}, \bibinfo
  {author} {\bibfnamefont {B.}~\bibnamefont {Shen}}, \bibinfo {author}
  {\bibfnamefont {Z.}~\bibnamefont {Xiao}}, \bibinfo {author} {\bibfnamefont
  {U.}~\bibnamefont {Welp}}, \ and\ \bibinfo {author} {\bibfnamefont
  {V.}~\bibnamefont {Vlasko-Vlasov}},\ }{\bibfield  {journal}
  {\bibinfo  {journal} {Physical Review B}\ }\textbf {\bibinfo {volume}
  {111}},\ \bibinfo {pages} {054439} (\bibinfo {year}
  {2025}{\natexlab{a}})}\BibitemShut {NoStop}%
\bibitem [{\citenamefont {Yin}\ \emph {et~al.}(2019)\citenamefont {Yin},
  \citenamefont {Zhang}, \citenamefont {Chang}, \citenamefont {Wang},
  \citenamefont {Tsirkin}, \citenamefont {Guguchia}, \citenamefont {Lian},
  \citenamefont {Zhou}, \citenamefont {Jiang}, \citenamefont {Belopolski} \emph
  {et~al.}}]{yin2019negative}%
  \BibitemOpen
  \bibfield  {author} {\bibinfo {author} {\bibfnamefont {J.-X.}\ \bibnamefont
  {Yin}}, \bibinfo {author} {\bibfnamefont {S.~S.}\ \bibnamefont {Zhang}},
  \bibinfo {author} {\bibfnamefont {G.}~\bibnamefont {Chang}}, \bibinfo
  {author} {\bibfnamefont {Q.}~\bibnamefont {Wang}}, \bibinfo {author}
  {\bibfnamefont {S.~S.}\ \bibnamefont {Tsirkin}}, \bibinfo {author}
  {\bibfnamefont {Z.}~\bibnamefont {Guguchia}}, \bibinfo {author}
  {\bibfnamefont {B.}~\bibnamefont {Lian}}, \bibinfo {author} {\bibfnamefont
  {H.}~\bibnamefont {Zhou}}, \bibinfo {author} {\bibfnamefont {K.}~\bibnamefont
  {Jiang}}, \bibinfo {author} {\bibfnamefont {I.}~\bibnamefont {Belopolski}},
  \emph {et~al.},\ }{\bibfield  {journal} {\bibinfo  {journal}
  {Nature Physics}\ }\textbf {\bibinfo {volume} {15}},\ \bibinfo {pages} {443}
  (\bibinfo {year} {2019})}\BibitemShut {NoStop}%
\bibitem [{\citenamefont {Nag}\ \emph {et~al.}(2022)\citenamefont {Nag},
  \citenamefont {Peng}, \citenamefont {Li}, \citenamefont {Agrestini},
  \citenamefont {Robarts}, \citenamefont {Garc{\'\i}a-Fern{\'a}ndez},
  \citenamefont {Walters}, \citenamefont {Wang}, \citenamefont {Yin},
  \citenamefont {Lei} \emph {et~al.}}]{nag2022correlation}%
  \BibitemOpen
  \bibfield  {author} {\bibinfo {author} {\bibfnamefont {A.}~\bibnamefont
  {Nag}}, \bibinfo {author} {\bibfnamefont {Y.}~\bibnamefont {Peng}}, \bibinfo
  {author} {\bibfnamefont {J.}~\bibnamefont {Li}}, \bibinfo {author}
  {\bibfnamefont {S.}~\bibnamefont {Agrestini}}, \bibinfo {author}
  {\bibfnamefont {H.}~\bibnamefont {Robarts}}, \bibinfo {author} {\bibfnamefont
  {M.}~\bibnamefont {Garc{\'\i}a-Fern{\'a}ndez}}, \bibinfo {author}
  {\bibfnamefont {A.}~\bibnamefont {Walters}}, \bibinfo {author} {\bibfnamefont
  {Q.}~\bibnamefont {Wang}}, \bibinfo {author} {\bibfnamefont {Q.}~\bibnamefont
  {Yin}}, \bibinfo {author} {\bibfnamefont {H.}~\bibnamefont {Lei}},  \emph
  {et~al.},\ }{\bibfield  {journal} {\bibinfo  {journal} {Nature
  Communications}\ }\textbf {\bibinfo {volume} {13}},\ \bibinfo {pages} {7317}
  (\bibinfo {year} {2022})}\BibitemShut {NoStop}%
\bibitem [{\citenamefont {Xu}\ \emph {et~al.}(2020)\citenamefont {Xu},
  \citenamefont {Zhao}, \citenamefont {Yi}, \citenamefont {Wang}, \citenamefont
  {Yin}, \citenamefont {Wang}, \citenamefont {Hu}, \citenamefont {Wang},
  \citenamefont {Liu}, \citenamefont {Xu} \emph {et~al.}}]{xu2020electronic}%
  \BibitemOpen
  \bibfield  {author} {\bibinfo {author} {\bibfnamefont {Y.}~\bibnamefont
  {Xu}}, \bibinfo {author} {\bibfnamefont {J.}~\bibnamefont {Zhao}}, \bibinfo
  {author} {\bibfnamefont {C.}~\bibnamefont {Yi}}, \bibinfo {author}
  {\bibfnamefont {Q.}~\bibnamefont {Wang}}, \bibinfo {author} {\bibfnamefont
  {Q.}~\bibnamefont {Yin}}, \bibinfo {author} {\bibfnamefont {Y.}~\bibnamefont
  {Wang}}, \bibinfo {author} {\bibfnamefont {X.}~\bibnamefont {Hu}}, \bibinfo
  {author} {\bibfnamefont {L.}~\bibnamefont {Wang}}, \bibinfo {author}
  {\bibfnamefont {E.}~\bibnamefont {Liu}}, \bibinfo {author} {\bibfnamefont
  {G.}~\bibnamefont {Xu}},  \emph {et~al.},\ }{\bibfield
  {journal} {\bibinfo  {journal} {Nature communications}\ }\textbf {\bibinfo
  {volume} {11}},\ \bibinfo {pages} {3985} (\bibinfo {year}
  {2020})}\BibitemShut {NoStop}%
\bibitem [{\citenamefont {Heide}\ \emph {et~al.}(2009)\citenamefont {Heide},
  \citenamefont {Bihlmayer},\ and\ \citenamefont
  {Bl{\"u}gel}}]{heide2009describing}%
  \BibitemOpen
  \bibfield  {author} {\bibinfo {author} {\bibfnamefont {M.}~\bibnamefont
  {Heide}}, \bibinfo {author} {\bibfnamefont {G.}~\bibnamefont {Bihlmayer}}, \
  and\ \bibinfo {author} {\bibfnamefont {S.}~\bibnamefont {Bl{\"u}gel}},\
  }{\bibfield  {journal} {\bibinfo  {journal} {Physica B:
  Condensed Matter}\ }\textbf {\bibinfo {volume} {404}},\ \bibinfo {pages}
  {2678} (\bibinfo {year} {2009})}\BibitemShut {NoStop}%
\bibitem [{\citenamefont {Kuz'{M}in}\ \emph {et~al.}(2020)\citenamefont
  {Kuz'{M}in}, \citenamefont {Skokov}, \citenamefont {Diop}, \citenamefont
  {Radulov},\ and\ \citenamefont {Gutfleisch}}]{kuz2020exchange}%
  \BibitemOpen
  \bibfield  {author} {\bibinfo {author} {\bibfnamefont {M.}~\bibnamefont
  {Kuz'{M}in}}, \bibinfo {author} {\bibfnamefont {K.}~\bibnamefont {Skokov}},
  \bibinfo {author} {\bibfnamefont {L.}~\bibnamefont {Diop}}, \bibinfo {author}
  {\bibfnamefont {I.}~\bibnamefont {Radulov}}, \ and\ \bibinfo {author}
  {\bibfnamefont {O.}~\bibnamefont {Gutfleisch}},\ }{\bibfield
  {journal} {\bibinfo  {journal} {The European Physical Journal Plus}\ }\textbf
  {\bibinfo {volume} {135}},\ \bibinfo {pages} {1} (\bibinfo {year}
  {2020})}\BibitemShut {NoStop}%
\bibitem [{\citenamefont {D{\"o}ring}(1957)}]{doring1957richtungsabhangigkeit}%
  \BibitemOpen
  \bibfield  {author} {\bibinfo {author} {\bibfnamefont {W.}~\bibnamefont
  {D{\"o}ring}},\ }{\bibfield  {journal} {\bibinfo  {journal}
  {Annalen der Physik}\ }\textbf {\bibinfo {volume} {456}},\ \bibinfo {pages}
  {102} (\bibinfo {year} {1957})}\BibitemShut {NoStop}%
\bibitem [{\citenamefont {Shen}\ \emph {et~al.}(2019)\citenamefont {Shen},
  \citenamefont {Zeng}, \citenamefont {Zhang}, \citenamefont {Tong},
  \citenamefont {Ling}, \citenamefont {Xi}, \citenamefont {Wang}, \citenamefont
  {Liu}, \citenamefont {Wang}, \citenamefont {Wu} \emph
  {et~al.}}]{shen2019anisotropies}%
  \BibitemOpen
  \bibfield  {author} {\bibinfo {author} {\bibfnamefont {J.}~\bibnamefont
  {Shen}}, \bibinfo {author} {\bibfnamefont {Q.}~\bibnamefont {Zeng}}, \bibinfo
  {author} {\bibfnamefont {S.}~\bibnamefont {Zhang}}, \bibinfo {author}
  {\bibfnamefont {W.}~\bibnamefont {Tong}}, \bibinfo {author} {\bibfnamefont
  {L.}~\bibnamefont {Ling}}, \bibinfo {author} {\bibfnamefont {C.}~\bibnamefont
  {Xi}}, \bibinfo {author} {\bibfnamefont {Z.}~\bibnamefont {Wang}}, \bibinfo
  {author} {\bibfnamefont {E.}~\bibnamefont {Liu}}, \bibinfo {author}
  {\bibfnamefont {W.}~\bibnamefont {Wang}}, \bibinfo {author} {\bibfnamefont
  {G.}~\bibnamefont {Wu}},  \emph {et~al.},\ }{\bibfield
  {journal} {\bibinfo  {journal} {Applied Physics Letters}\ }\textbf {\bibinfo
  {volume} {115}} (\bibinfo {year} {2019})}\BibitemShut {NoStop}%
\bibitem [{\citenamefont {Lee}\ \emph {et~al.}(2022)\citenamefont {Lee},
  \citenamefont {Vir}, \citenamefont {Manna}, \citenamefont {Shekhar},
  \citenamefont {Moore}, \citenamefont {Kastner}, \citenamefont {Felser},\ and\
  \citenamefont {Orenstein}}]{lee2022observation}%
  \BibitemOpen
  \bibfield  {author} {\bibinfo {author} {\bibfnamefont {C.}~\bibnamefont
  {Lee}}, \bibinfo {author} {\bibfnamefont {P.}~\bibnamefont {Vir}}, \bibinfo
  {author} {\bibfnamefont {K.}~\bibnamefont {Manna}}, \bibinfo {author}
  {\bibfnamefont {C.}~\bibnamefont {Shekhar}}, \bibinfo {author} {\bibfnamefont
  {J.}~\bibnamefont {Moore}}, \bibinfo {author} {\bibfnamefont
  {M.}~\bibnamefont {Kastner}}, \bibinfo {author} {\bibfnamefont
  {C.}~\bibnamefont {Felser}}, \ and\ \bibinfo {author} {\bibfnamefont
  {J.}~\bibnamefont {Orenstein}},\ }{\bibfield  {journal}
  {\bibinfo  {journal} {Nature communications}\ }\textbf {\bibinfo {volume}
  {13}},\ \bibinfo {pages} {3000} (\bibinfo {year} {2022})}\BibitemShut
  {NoStop}%
\bibitem [{\citenamefont {Shiogai}\ \emph {et~al.}(2022)\citenamefont
  {Shiogai}, \citenamefont {Ikeda}, \citenamefont {Fujiwara}, \citenamefont
  {Seki}, \citenamefont {Takanashi},\ and\ \citenamefont
  {Tsukazaki}}]{shiogai2022electrical}%
  \BibitemOpen
  \bibfield  {author} {\bibinfo {author} {\bibfnamefont {J.}~\bibnamefont
  {Shiogai}}, \bibinfo {author} {\bibfnamefont {J.}~\bibnamefont {Ikeda}},
  \bibinfo {author} {\bibfnamefont {K.}~\bibnamefont {Fujiwara}}, \bibinfo
  {author} {\bibfnamefont {T.}~\bibnamefont {Seki}}, \bibinfo {author}
  {\bibfnamefont {K.}~\bibnamefont {Takanashi}}, \ and\ \bibinfo {author}
  {\bibfnamefont {A.}~\bibnamefont {Tsukazaki}},\ }{\bibfield
  {journal} {\bibinfo  {journal} {Physical Review Materials}\ }\textbf
  {\bibinfo {volume} {6}},\ \bibinfo {pages} {114203} (\bibinfo {year}
  {2022})}\BibitemShut {NoStop}%
\bibitem [{\citenamefont {Xu}\ \emph {et~al.}(2018)\citenamefont {Xu},
  \citenamefont {Liu}, \citenamefont {Shi}, \citenamefont {Muechler},
  \citenamefont {Gayles}, \citenamefont {Felser},\ and\ \citenamefont
  {Sun}}]{xu2018topological}%
  \BibitemOpen
  \bibfield  {author} {\bibinfo {author} {\bibfnamefont {Q.}~\bibnamefont
  {Xu}}, \bibinfo {author} {\bibfnamefont {E.}~\bibnamefont {Liu}}, \bibinfo
  {author} {\bibfnamefont {W.}~\bibnamefont {Shi}}, \bibinfo {author}
  {\bibfnamefont {L.}~\bibnamefont {Muechler}}, \bibinfo {author}
  {\bibfnamefont {J.}~\bibnamefont {Gayles}}, \bibinfo {author} {\bibfnamefont
  {C.}~\bibnamefont {Felser}}, \ and\ \bibinfo {author} {\bibfnamefont
  {Y.}~\bibnamefont {Sun}},\ }{\bibfield  {journal} {\bibinfo
  {journal} {Physical Review B}\ }\textbf {\bibinfo {volume} {97}},\ \bibinfo
  {pages} {235416} (\bibinfo {year} {2018})}\BibitemShut {NoStop}%
\bibitem [{\citenamefont {Gayles}\ \emph {et~al.}(2015)\citenamefont {Gayles},
  \citenamefont {Freimuth}, \citenamefont {Schena}, \citenamefont {Lani},
  \citenamefont {Mavropoulos}, \citenamefont {Duine}, \citenamefont
  {Bl{\"u}gel}, \citenamefont {Sinova},\ and\ \citenamefont
  {Mokrousov}}]{gayles2015dzyaloshinskii}%
  \BibitemOpen
  \bibfield  {author} {\bibinfo {author} {\bibfnamefont {J.}~\bibnamefont
  {Gayles}}, \bibinfo {author} {\bibfnamefont {F.}~\bibnamefont {Freimuth}},
  \bibinfo {author} {\bibfnamefont {T.}~\bibnamefont {Schena}}, \bibinfo
  {author} {\bibfnamefont {G.}~\bibnamefont {Lani}}, \bibinfo {author}
  {\bibfnamefont {P.}~\bibnamefont {Mavropoulos}}, \bibinfo {author}
  {\bibfnamefont {R.}~\bibnamefont {Duine}}, \bibinfo {author} {\bibfnamefont
  {S.}~\bibnamefont {Bl{\"u}gel}}, \bibinfo {author} {\bibfnamefont
  {J.}~\bibnamefont {Sinova}}, \ and\ \bibinfo {author} {\bibfnamefont
  {Y.}~\bibnamefont {Mokrousov}},\ }{\bibfield  {journal}
  {\bibinfo  {journal} {Physical review letters}\ }\textbf {\bibinfo {volume}
  {115}},\ \bibinfo {pages} {036602} (\bibinfo {year} {2015})}\BibitemShut
  {NoStop}%
\bibitem [{\citenamefont {Zyuzin}\ and\ \citenamefont
  {Tiwari}(2016)}]{zyuzin2016intrinsic}%
  \BibitemOpen
  \bibfield  {author} {\bibinfo {author} {\bibfnamefont {A.~A.}\ \bibnamefont
  {Zyuzin}}\ and\ \bibinfo {author} {\bibfnamefont {R.~P.}\ \bibnamefont
  {Tiwari}},\ }{\bibfield  {journal} {\bibinfo  {journal} {JETP
  letters}\ }\textbf {\bibinfo {volume} {103}},\ \bibinfo {pages} {717}
  (\bibinfo {year} {2016})}\BibitemShut {NoStop}%
\bibitem [{\citenamefont {Burkov}\ and\ \citenamefont
  {Balents}(2011)}]{burkov2011weyl}%
  \BibitemOpen
  \bibfield  {author} {\bibinfo {author} {\bibfnamefont {A.}~\bibnamefont
  {Burkov}}\ and\ \bibinfo {author} {\bibfnamefont {L.}~\bibnamefont
  {Balents}},\ }{\bibfield  {journal} {\bibinfo  {journal}
  {Physical review letters}\ }\textbf {\bibinfo {volume} {107}},\ \bibinfo
  {pages} {127205} (\bibinfo {year} {2011})}\BibitemShut {NoStop}%
\bibitem [{\citenamefont {Das}\ and\ \citenamefont
  {Agarwal}(2019)}]{das2019linear}%
  \BibitemOpen
  \bibfield  {author} {\bibinfo {author} {\bibfnamefont {K.}~\bibnamefont
  {Das}}\ and\ \bibinfo {author} {\bibfnamefont {A.}~\bibnamefont {Agarwal}},\
  }{\bibfield  {journal} {\bibinfo  {journal} {Physical Review
  B}\ }\textbf {\bibinfo {volume} {99}},\ \bibinfo {pages} {085405} (\bibinfo
  {year} {2019})}\BibitemShut {NoStop}%
\bibitem [{\citenamefont {Wei}\ \emph {et~al.}(2022)\citenamefont {Wei},
  \citenamefont {Yang}, \citenamefont {Shen},\ and\ \citenamefont
  {Tao}}]{wei2022ferromagnetic}%
  \BibitemOpen
  \bibfield  {author} {\bibinfo {author} {\bibfnamefont {X.-P.}\ \bibnamefont
  {Wei}}, \bibinfo {author} {\bibfnamefont {N.}~\bibnamefont {Yang}}, \bibinfo
  {author} {\bibfnamefont {J.}~\bibnamefont {Shen}}, \ and\ \bibinfo {author}
  {\bibfnamefont {X.}~\bibnamefont {Tao}},\ }{\bibfield
  {journal} {\bibinfo  {journal} {Physica E: Low-dimensional Systems and
  Nanostructures}\ }\textbf {\bibinfo {volume} {140}},\ \bibinfo {pages}
  {115164} (\bibinfo {year} {2022})}\BibitemShut {NoStop}%
\bibitem [{\citenamefont {Zhang}\ \emph {et~al.}(2017)\citenamefont {Zhang},
  \citenamefont {Xu}, \citenamefont {Wang}, \citenamefont {Lin}, \citenamefont
  {Du}, \citenamefont {Guo}, \citenamefont {Lee}, \citenamefont {Lu},
  \citenamefont {Feng}, \citenamefont {Huang} \emph
  {et~al.}}]{zhang2017magnetic}%
  \BibitemOpen
  \bibfield  {author} {\bibinfo {author} {\bibfnamefont {C.-L.}\ \bibnamefont
  {Zhang}}, \bibinfo {author} {\bibfnamefont {S.-Y.}\ \bibnamefont {Xu}},
  \bibinfo {author} {\bibfnamefont {C.}~\bibnamefont {Wang}}, \bibinfo {author}
  {\bibfnamefont {Z.}~\bibnamefont {Lin}}, \bibinfo {author} {\bibfnamefont
  {Z.}~\bibnamefont {Du}}, \bibinfo {author} {\bibfnamefont {C.}~\bibnamefont
  {Guo}}, \bibinfo {author} {\bibfnamefont {C.-C.}\ \bibnamefont {Lee}},
  \bibinfo {author} {\bibfnamefont {H.}~\bibnamefont {Lu}}, \bibinfo {author}
  {\bibfnamefont {Y.}~\bibnamefont {Feng}}, \bibinfo {author} {\bibfnamefont
  {S.-M.}\ \bibnamefont {Huang}},  \emph {et~al.},\ }{\bibfield
  {journal} {\bibinfo  {journal} {Nature Physics}\ }\textbf {\bibinfo {volume}
  {13}},\ \bibinfo {pages} {979} (\bibinfo {year} {2017})}\BibitemShut
  {NoStop}%
\bibitem [{\citenamefont {Pate}\ \emph
  {et~al.}(2025{\natexlab{b}})\citenamefont {Pate}, \citenamefont {Wang},
  \citenamefont {Shen}, \citenamefont {Xiao}, \citenamefont {Welp},\ and\
  \citenamefont {Vlasko-Vlasov}}]{2025anomalies_inplane_hall}%
  \BibitemOpen
  \bibfield  {author} {\bibinfo {author} {\bibfnamefont {S.}~\bibnamefont
  {Pate}}, \bibinfo {author} {\bibfnamefont {B.}~\bibnamefont {Wang}}, \bibinfo
  {author} {\bibfnamefont {B.}~\bibnamefont {Shen}}, \bibinfo {author}
  {\bibfnamefont {Z.}~\bibnamefont {Xiao}}, \bibinfo {author} {\bibfnamefont
  {U.}~\bibnamefont {Welp}}, \ and\ \bibinfo {author} {\bibfnamefont
  {V.}~\bibnamefont {Vlasko-Vlasov}},\ }{\bibfield  {journal}
  {\bibinfo  {journal} {Physical Review B}\ }\textbf {\bibinfo {volume}
  {111}},\ \bibinfo {pages} {054439} (\bibinfo {year}
  {2025}{\natexlab{b}})}\BibitemShut {NoStop}%
\bibitem{fleurWeb}
{The FLEUR project}, \url{https://www.flapw.de/}.%
\bibitem [{\citenamefont {Wimmer}\ \emph {et~al.}(1981)\citenamefont {Wimmer},
  \citenamefont {Krakauer}, \citenamefont {Weinert},\ and\ \citenamefont
  {Freeman}}]{FLAPW}%
  \BibitemOpen
  \bibfield  {author} {\bibinfo {author} {\bibfnamefont {E.}~\bibnamefont
  {Wimmer}}, \bibinfo {author} {\bibfnamefont {H.}~\bibnamefont {Krakauer}},
  \bibinfo {author} {\bibfnamefont {M.}~\bibnamefont {Weinert}}, \ and\
  \bibinfo {author} {\bibfnamefont {A.~J.}\ \bibnamefont {Freeman}},\ }\href
  {\doibase 10.1103/PhysRevB.24.864} {\bibfield  {journal} {\bibinfo  {journal}
  {Phys. Rev. B}\ }\textbf {\bibinfo {volume} {24}},\ \bibinfo {pages} {864}
  (\bibinfo {year} {1981})}\BibitemShut {NoStop}%
\bibitem [{\citenamefont {Perdew}\ \emph {et~al.}(1996)\citenamefont {Perdew},
  \citenamefont {Burke},\ and\ \citenamefont
  {Ernzerhof}}]{perdew1996generalized}%
  \BibitemOpen
  \bibfield  {author} {\bibinfo {author} {\bibfnamefont {J.~P.}\ \bibnamefont
  {Perdew}}, \bibinfo {author} {\bibfnamefont {K.}~\bibnamefont {Burke}}, \
  and\ \bibinfo {author} {\bibfnamefont {M.}~\bibnamefont {Ernzerhof}},\
  }{\bibfield  {journal} {\bibinfo  {journal} {Physical review
  letters}\ }\textbf {\bibinfo {volume} {77}},\ \bibinfo {pages} {3865}
  (\bibinfo {year} {1996})}\BibitemShut {NoStop}%
\bibitem [{\citenamefont {Weihrich}\ \emph {et~al.}(2015)\citenamefont
  {Weihrich}, \citenamefont {Yan}, \citenamefont {Rothballer}, \citenamefont
  {Peter}, \citenamefont {Rommel}, \citenamefont {Haumann}, \citenamefont
  {Winter}, \citenamefont {Schwickert},\ and\ \citenamefont
  {P{\"o}ttgen}}]{weihrich2015tuneable}%
  \BibitemOpen
  \bibfield  {author} {\bibinfo {author} {\bibfnamefont {R.}~\bibnamefont
  {Weihrich}}, \bibinfo {author} {\bibfnamefont {W.}~\bibnamefont {Yan}},
  \bibinfo {author} {\bibfnamefont {J.}~\bibnamefont {Rothballer}}, \bibinfo
  {author} {\bibfnamefont {P.}~\bibnamefont {Peter}}, \bibinfo {author}
  {\bibfnamefont {S.~M.}\ \bibnamefont {Rommel}}, \bibinfo {author}
  {\bibfnamefont {S.}~\bibnamefont {Haumann}}, \bibinfo {author} {\bibfnamefont
  {F.}~\bibnamefont {Winter}}, \bibinfo {author} {\bibfnamefont
  {C.}~\bibnamefont {Schwickert}}, \ and\ \bibinfo {author} {\bibfnamefont
  {R.}~\bibnamefont {P{\"o}ttgen}},\ }{\bibfield  {journal}
  {\bibinfo  {journal} {Dalton Transactions}\ }\textbf {\bibinfo {volume}
  {44}},\ \bibinfo {pages} {15855} (\bibinfo {year} {2015})}\BibitemShut
  {NoStop}%
\end{thebibliography}
\end{document}